\documentclass[aps,prl,
showpacs,
superscriptaddress,reprint,floatfix,citeautoscript,twocolumn]{revtex4-2}
\usepackage{graphicx}
\newcommand{\RNum}[1]{\uppercase\expandafter{\romannumeral #1\relax}}
\usepackage{amsmath}
\usepackage{amssymb}
\usepackage{bm}
\usepackage{float}
\usepackage{dcolumn}
\usepackage{subfigure}
\usepackage{hyperref}
\usepackage[usenames]{color}
\usepackage[dvipsnames]{xcolor}
\usepackage{booktabs}
\usepackage{multirow}
\usepackage{siunitx}
\usepackage{wrapfig}
\usepackage{lipsum}
\usepackage{setspace}
\usepackage{xcolor}
\usepackage{longtable}
\hypersetup{pdfborder=0 0 0,colorlinks=true,citecolor=blue,linkcolor=blue}
\input epsf

\usepackage{ulem}

\begin{document}

\title{Terahertz Slonczewski propagating spin waves {\color{black} and large output voltage} in antiferromagnetic spin-Hall nano-oscillators}
\author{M. Hamdi}
\email[]{mohammad.hamdi@epfl.ch, \\
	mohamad.hamdi90@gmail.com}
\affiliation{\'Ecole Polytechnique F\'ed\'erale de Lausanne (EPFL),  Institute of Materials, Laboratory of Nanoscale Magnetic Materials and Magnonics,  CH-1015 Lausanne, Switzerland}

\author{D.~Grundler}
\email[]{dirk.grundler@epfl.ch}
\affiliation{\'Ecole Polytechnique F\'ed\'erale de Lausanne (EPFL),  Institute of Materials, Laboratory of Nanoscale Magnetic Materials and Magnonics,  CH-1015 Lausanne, Switzerland}
\affiliation{\'Ecole Polytechnique F\'ed\'erale de Lausanne (EPFL), Institute of Electrical and Micro engineering, CH-1015 Lausanne, Switzerland}

\date{\today}

\begin{abstract}
 We study theoretically antiferromagnet (AFM) based spin-Hall nano-oscillators (SHNOs) consisting of a nano-constriction (NC) in a thin-film uniaxial AFM. {\color{black}By solving the derived SW equation we evidence radially propagating spin waves (SWs) at THz frequencies similar to the Slonczewski SWs known at GHz frequencies for a ferromagnet-based SHNO. We predict a minimum threshold current for a specific NC radius accessible by the state-of-the-art nanotechnology. The exchange interaction enhanced spin pumping for AFMs leads to a strong thickness dependent threshold frequency. We show that the uniaxial AFMs generate ac electrical fields via spin pumping that are three orders of magnitude larger than reported for biaxial AFMs. Our work enhances the fundamental understanding of current-driven SWs in AFM-SHNOs and enables optimization of practical devices in terms of material choice, device geometry, and frequency tunability. The propagating SWs offer remote THz signal generation and an efficient means for synchronization of SHNOs when aiming at high power.}


\end{abstract}

\maketitle



\begin{figure}[b]
	\centering
	\includegraphics[scale=0.29]{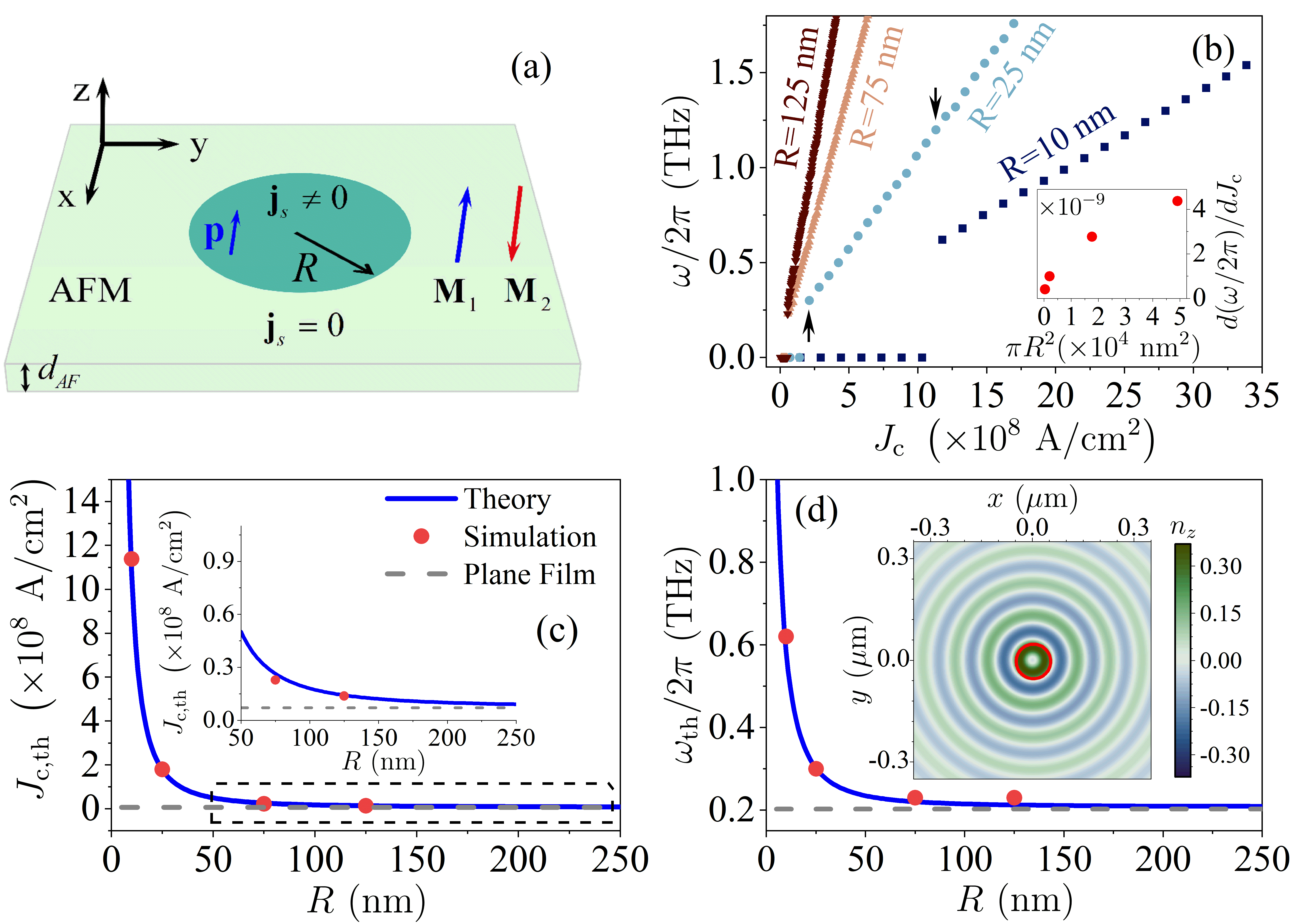}
	\caption{(a) Schematics of the studied system. The spin current with polarization $\mathbf{p}$ originates from the SHE of the electric current constrained in a circular NC of radius $R$ in the adjacent HM layer. The spin polarization of the spin current is assumed to be along the N\'eel order vector. (b) Excitation frequency versus applied current density  for the simulated SHNOs with different NC radii $R$. {\color{black} The black upward (downward) arrow indicates the mode for $R=25$ nm at threshold $J_{c,\text{th}}=2.12\times 10^8$ A/cm$^{2}$ with $f_{\text{th}}=0.3$ THz ($J_{c,\text{h}}=11.3\times 10^8$ A/cm$^{2}$ with $f_{\text{h}}=1.2$ THz) which is selected to calculate output electric fields due to spin pumping in Fig. \ref{figure4}.} The slope $\beta$ as a function of $R^2$ is shown in the inset. Threshold current density (c) and frequency (d) obtained from simulations (filled red circles) and Eqs. \ref{ThresholdCurrentDensity} and \ref{ThresholdFrequency} (blue line), respectively. Gray dashed lines indicate the plane film limit. Inset in (d) shows a snapshot of current induced SW at threshold for SHNO with $R=50$ nm (red circle) and simulation parameters given in the supplementary \cite{SupplementalNote1}.}
	\label{figure1}
\end{figure}

\textit{Introduction.}--- Terahertz (THz) radiation attracts tremendous attention due to its several promising applications including ultrafast communication \cite{Nagatsuma2016,Ma2019}, biosensing and imaging \cite{Sirtori2002,Osborne2008}. However, the widespread use of THz instrumentation is constrained, in particular, because of the lack of efficient sources and detectors of THz radiation for frequencies in the range of 0.1-10 THz (THz gap) \cite{Sirtori2002,Osborne2008}. Therefore, there is an urgent need to develop small-sized solid-state THz sources and detectors. Spin wave (SW) excitations in antiferromagnetic materials (AFMs) offer frequencies which cover perfectly the THz gap. These materials can be utilized to close the THz gap in terms of antiferromagnetic spintronic and magnonic devices \cite{Cheng2014,Jungwirth2016,Baltz2018,Jungwirth2018,Duine2018,Gomonay2018a,Zelezny2018,Nemec2018,Smejkal2018,Cheng2016,Khymyn2017a,Sulymenko2017,Johansen2017,Checinski2017,Khymyn2017,Johansen2018,Gomonay2018,Puliafito2019,Lee2019,Lisenkov2019,Troncoso2019,Safin2020,Parthasarathy2021,Shtanko2021,Sulymenko2018,Khymyn2018}. \\
\indent One of the most promising devices is the AFM-based spin-Hall nano-oscillator (AFM-SHNO) consisting of an AFM and a heavy metal (HM) layer \cite{Cheng2016,Khymyn2017a,Sulymenko2017,Johansen2017,Checinski2017,Khymyn2017,Johansen2018,Gomonay2018,Puliafito2019,Lee2019,Lisenkov2019,Troncoso2019,Safin2020,Parthasarathy2021,Shtanko2021,Sulymenko2018,Khymyn2018}. In an AFM-SHNO, a spin current generated by the spin-Hall effect (SHE) in the HM enters the AFM layer and exerts the so-called spin-orbit torque (SOT) \cite{stiles2002,ralph2008}. It has been shown theoretically that when the spin current exceeds a certain threshold, damping can {\color{black} be compensated}, and the magnetization starts self-sustained auto-oscillations. They can generate THz radiation via the combination of spin pumping \cite{Cheng2014,Johansen2017,Li2020,Vaidya2020} and inverse spin-Hall effect \cite{Cheng2016,Khymyn2017a,Sulymenko2017,Johansen2017,Checinski2017,Khymyn2017,Johansen2018,Gomonay2018,Puliafito2019,Lee2019,Lisenkov2019,Troncoso2019,Safin2020,Parthasarathy2021,Shtanko2021,Sulymenko2018,Khymyn2018}. The AFM-SHNOs have potential not only as THz sources and detectors \cite{Cheng2016,Khymyn2017a,Sulymenko2017,Johansen2017,Checinski2017,Khymyn2017,Johansen2018,Gomonay2018,Puliafito2019,Lee2019,Lisenkov2019,Troncoso2019,Safin2020,Parthasarathy2021,Shtanko2021}, but also as ultra-fast artificial neurons which generate picosecond-duration spikes \cite{Sulymenko2018,Khymyn2018} and enable ultra-fast neuromorphic applications \cite{Daniels2018,Sulymenko2019,Sulymenko2019a,Grollier2020}. Howver, there has not yet been any report on the experimental realization of an AFM-SHNOs. \\
\indent All the theoretical studies mentioned above were carried out by considering an AFM in the macrospin approximation. They described uniform antiferromagnetic resonance (AFMR) excitation by a spin current in planar bilayer HM/AFM structures \cite{Cheng2016,Khymyn2017a,Sulymenko2017,Johansen2017,Checinski2017,Khymyn2017,Johansen2018,Gomonay2018,Puliafito2019,Lee2019,Lisenkov2019,Troncoso2019,Safin2020,Parthasarathy2021,Shtanko2021,Sulymenko2018,Khymyn2018}. Still, there have been few micromagnetic studies of SHNOs addressing the planar HM/AFM geometry and AFMR excitation \cite{Puliafito2019,Lee2019}. These studies do not describe however a realistic AFM-SHNO which incorporates a nano-constriction (NC) \cite{Chen2016,Demidov2017,Hoefer2010,Mohseni2013,Mohseni2018}. {\color{black} NCs are essential for ferromagnet-based SHNOs (FM-SHNOs) which excite propagating SWs \cite{slonczewski1996,Slonczewski1999,Hoefer2005,Mohseni2018b}. These SWs enable the efficient synchronization of FM-SHNOs \cite{Kendziorczyk2014,Houshang2016,Kendziorczyk2016,Awad2017,Zahedinejad2020,Zahedinejad2022} granting sufficient signal quality for applications.}\\
\indent In this Letter, we study theoretically and by means of micromagnetic simulations a NC-based AFM-SHNO and provide a fundamental understanding of its spin dynamics. We first derive the SW equation for an insulating AFM layer subjected to a spin current in a circular NC. We show that the solution to this equation is the antiferromagnetic counterpart of Slonczewski SWs known for ferromagnetic SHNOs. The Slonczewski spin waves are radially propagating waves generated by spin torque in a circular NC in a FM layer \cite{slonczewski1996,berger1996,Slonczewski1999,Hoefer2005,Chung2016}. We obtain the threshold frequency and current for such THz antiferromagnetic Slonczewski SWs (THz-ASSWs) as a function of the material parameters and the NC radius $R$. Our analytical results show good agreement with the micromagnetic simulations. A unique feature of these THz-ASSWs is that the threshold current exhibits a minimum with respect to the NC radius which allows one to reduce the power consumption to a minimum. Interestingly, the minimum diameter $2R$ is in the range of few tens to few hundreds of nanometers for realistic material parameters and is realizable by state-of-the-art fabrication technology. {\color{black} Moreover, we show that when used in NC-SHNOs, the uniaxial AFMs generate THz outputs three orders of magnitude larger compared to uniformly excited biaxial AFM-SHNOs. This is a significant improvement on both the power consumption and output signal of the AFM-SHNOs and draws the attention to the previously abandoned uniaxial AFMs which were considered not suitable for THz SHNOs \cite{Cheng2016,Khymyn2017a}.	
}\\
\indent \textit{Theory.}--- We consider an AFM-SHNO as a plane AFM film that is subjected to a spin current injected locally in a circular region of radius $R$, as depicted in Fig. \ref{figure1} (a). Such spin current is generated by restricting the electric current in the adjacent HM into a NC \cite{Chen2016,Demidov2017}. We assume an AFM with uniaxial anisotropy in the plane along the $x$-axis and a thickness, $d_{AF}$, smaller than the exchange length, $\lambda_{ex}=\sqrt{a^{2}\omega_{E}/4\omega_{A}}$. Thereby we avoid standing waves along the thickness. Here, $a$ is the magnetic lattice constant of the material and $\omega_{E}=\gamma H_E$ ($\omega_{A}=\gamma H_A$) is the exchange (anisotropy) field in the units of frequency with $H_E$ ($H_A$) being the exchange (anisotropy) field and $\gamma$ is the electron gyromagnetic ratio. The N\'eel order vector and average magnetization are defined as $\mathbf{n}=(\mathbf{M}_{1}-\mathbf{M}_{2})/2M_{s}$ and $\mathbf{m}=(\mathbf{M}_{1}+\mathbf{M}_{2})/2M_{s}$, respectively, where $\mathbf{n}.\mathbf{m}=0$ and $|\mathbf{n}|^2+|\mathbf{m}|^2=1$. $\mathbf{M}_{i}$ ($i=1,2$) is the sublattice magnetization with $|\mathbf{M}_{i}|=M_s$. We consider the polarization direction of the spin current, $\mathbf{\hat{p}}||\mathbf{n}$, where the SOT has the highest efficiency \cite{Lee2019}. There is no externally applied magnetic field. The total magnetic free energy of the AFM is $F=\int fdV$, where the energy density, $f$, is given by $f=\frac{M_{s}}{\gamma }\left\{\vphantom{\int}\right. \frac{1}{2}\omega_{E}\left( \mathbf{m}^{2}-\mathbf{n}^{2}\right) -\frac{1}{2}\omega_{A}\left(m_x^2 + n_x^2\right)-\frac{1}{2}\omega _{A}\lambda _{ex}^{2}\left[ \left( \mathbf{\nabla }\mathbf{m}\right)^{2}-\left( \mathbf{\nabla }\mathbf{n}\right)^{2}\right] \left. \vphantom{\int} \right\}$ \cite{Hals2011,Cheng2014,Galkina2018,Johansen2017,Wang2021}.
\indent The dynamics of $\mathbf{m}$ and $\mathbf{n}$ is governed by the Landau-Lifshitz-Gilbert-Slonczewski (LLGS) equations as
\begin{subequations}\label{LLGS}
\begin{eqnarray}
\mathbf{\dot{m}}=\bm{\omega }_{m}\times \mathbf{m}+\bm{\omega }_{n}\times \mathbf{n}+\alpha \left( \mathbf{m}\times \mathbf{\dot{m}}+\mathbf{n}\times \mathbf{\dot{n}}\right) +\bm{\tau }_{m}^{\text{s}}, \quad \label{LLGSm}
\\
\mathbf{\dot{n}}= \bm{\omega }_{m}\times \mathbf{n}+\bm{\omega }_{n}\times \mathbf{m} +\alpha\left( \mathbf{m}\times \mathbf{\dot{n}}+\mathbf{n}\times \mathbf{\dot{m}}\right)+\bm{\tau }_{n}^{\text{s}}, \quad \label{LLGSn}
\end{eqnarray}
\end{subequations}
where $\bm{\omega }_{m}=-\frac{\gamma }{M_{s}}\frac{\delta F}{\delta \mathbf{m}}$ and $\bm{\omega }_{n}=-\frac{\gamma }{M_{s}}\frac{\delta F}{\delta \mathbf{n}}$ are the effective fields in frequency units \cite{Hals2011,Cheng2014,Johansen2017,Wang2021}. $\alpha=\alpha_0 +\alpha_{\text{SP}}$ is the effective damping consisting of the Gilbert damping, $\alpha_0$, and the enhanced damping due to spin pumping, $\alpha_{\text{SP}}$ \cite{Johansen2017,Cheng2014}. The SOT terms are given as $\bm{\tau }_{m}^{\text{s}}=\omega _{s} \left[ \vphantom{\int} \right. \mathbf{m}\times \left( \mathbf{m}\times \mathbf{\hat{p}}\right) +\mathbf{n}\times \left( \mathbf{n}\times \mathbf{\hat{p}}\right) \left. \vphantom{\int} \right] H\left( R-r\right)$ and $\bm{\tau }_{n}^{\text{s}}=\omega _{s}\left[ \vphantom{\int} \right. \mathbf{m}\times \left( \mathbf{n}\times \mathbf{\hat{p}}\right) +\mathbf{n}\times \left( \mathbf{m}\times \mathbf{\hat{p}}\right) \left. \vphantom{\int} \right] H\left( R-r\right) $ where $\omega_{s}$ is the SOT strength in frequency units, $H(R-r)$ is the Heaviside step function and $r$ is the radial distance from the center of the NC. We consider the ansatz  $\mathbf{n}=n_{0}\mathbf{\hat{x}}+\mathbf{n}_{\bot}\left( x,y\right) e^{i\omega t}$ with $|\mathbf{n}_{\bot}\left( x,y\right)|\ll 1$ to derive the wave equation for small amplitude SWs excited by SOT. In the exchange limit ($\omega _{A}/\omega _{E}\ll1$), which is the case for most antiferromagnets and the absence of an external field, $\mathbf{m}\simeq (\omega _{A}/\omega _{E})\mathbf{n}_{\bot }\ll \mathbf{n}_{\bot }$ and $\left| n_{0}\right| \simeq 1$. By keeping the linear terms in $\mathbf{m}$ and $\mathbf{n}_{\bot}$ and eliminating $\mathbf{m}$ between the Eqs. \ref{LLGSm} and \ref{LLGSn} \cite{SupplementalNote1}, we obtain the wave equation for complex SW amplitude, $u=n_{y}+in_{z}$. By solving this equation we derive the threshold frequency and current density for the excited mode as
\begin{widetext}
\begin{eqnarray}
&&\omega_{\text{th}} =\dfrac{\alpha A_{12}\left( 2\omega _{E}+\omega _{A}\right)}{2\left( 1-\alpha A_{12}\right) } + \sqrt{\left[ \dfrac{\alpha A_{12}\left( 2\omega _{E}+\omega _{A}\right)}{2\left( 1-\alpha A_{12}\right) } \right]^{2}+ \left( 2\omega _{E}+\omega _{A}\right) \left[ \omega _{A}+\left( \dfrac{A_{0}+\alpha A_{11}}{1-\alpha A_{12}}\right) \left( \dfrac{\omega _{A}\lambda _{ex}^{2}}{R^{2}}\right) \right]},  \label{ThresholdFrequency}
\\
&&J_{c,\text{th}}=\left( \dfrac{eM_{s}d_{AF}}{2\gamma \hbar \theta_{SH}\eta}\right) \left\{\left( \dfrac{\alpha B_{12}}{1-\alpha A_{12}}\right) \omega_{\text{th}} +\left( \dfrac{\omega_{A}\lambda_{ex}^{2}}{R^{2}}\right) \left[ B_{0}+\alpha B_{11}+\alpha B_{12}\left( \dfrac{A_{0}+\alpha A_{11}}{1-\alpha A_{12}}\right) \right] \right\}, \label{ThresholdCurrentDensity}
\end{eqnarray}
\end{widetext}
respectively, where $A_{0}=1.428$, $A_{11}=-3.087$, $A_{12}=0.815$, $B_0=1.863$, $B_{11}=-1.581$ and $B_{12}=1.004$. \\

\indent \textit{Results and discussion.}--- By taking the fast-Fourier-transform (FFT) of $\mathbf{n}$, obtained from micromagnetic simulations \cite{SupplementalNote1,Donahue1999,Venkat2018}, we extract the power spectral density (PSD) of the magnetization dynamics.
Varying the current density and extracting the peak position of the PSD, we obtained the frequency versus current density curve of the SHNOs plotted in Fig. \ref{figure1}(b). {\color{black} The frequency variation for large currents is attributed to the excessive SOT beyond threshold according to $\omega=\omega_{\text{th}}+\beta (J_{c}-J_{\text{th}})$. The inset shows the SW frequency shift coefficient $\beta=d\omega/dJ_{c}$ as a function of the NC area. $\beta$ as determined from micromagnetic simulations increases with $R$ and seems to asymptotically approach the macrospin limit. The observed tunability of $\beta$ goes beyond a constant value in macrospin studies and can be used to engineer NC performance.} \newline
\indent Now we focus on the dynamics at the threshold, where the PSD peak jumps from zero to the threshold frequency. The threshold current density and corresponding frequency decrease with the NC radius as plotted in Fig. \ref{figure1}(c) and (d), respectively. {\color{black}The values $J_{c,\text{th}}$ obtained from micromagnetic simulations (red symbols) are plotted with an offset of $\Delta J=-0.4\times 10^8$ A/cm$^{2}$. We attribute this systematic error to the Taylor expansion of the complex transcendental equation that we employed to solve the eigenvalue problem \cite{SupplementalNote1}.  This fixed offset does not affect the main conclusion of the comparison, i.e., the significant dependency of $J_{c,\text{th}}$ on $R$ making a geometrical optimization of a NC device necessary. The inset in Fig. \ref{figure1}(c) shows a zoom in of the region indicated by dashed rectangle.}\\ 
\indent For large $R\gg\lambda_{ex}$ the threshold current density $J_{c,\text{th}}$ (frequency $\omega_{\rm th}$) asymptotically approaches the AFMR limit in a thin film, $J_{c,\text{th}}=\left( \frac{eM_{s}d_{AF}}{2\gamma \hbar \theta_{SH}\eta}\right)\alpha\omega_{0}$  ($\omega_{0}=\sqrt{\left( 2\omega _{E}+\omega _{A}\right)\omega _{A}}$), indicated by the gray dashed line in Fig. \ref{figure1}(c) (\ref{figure1}(d)). The inset in Fig. \ref{figure1}(d) shows a snapshot of the spin-current-generated spin wave simulated for $R=50~$nm. For small $R$ (below almost $\lambda_{ex}= 22.3$ nm), the exchange term dominates and the threshold current density and frequency are large. One can understand the increase in threshold current density with decreasing $R$ by considering the energy flow away from the NC by outgoing SWs. The wave vector, $k$, of the excited spin wave at the threshold is comparable to $2\pi/R$. Therefore, by decreasing $R$ one increases $k$ and the group velocity of the excited wave (following the dispersion in Eq. \ref{ThresholdFrequency}) and hence, the propagating energy away from the NC. To sustain this energy flow, a higher threshold current is necessary.
\begin{figure}[b]
	\centering
	\includegraphics[scale=0.285]{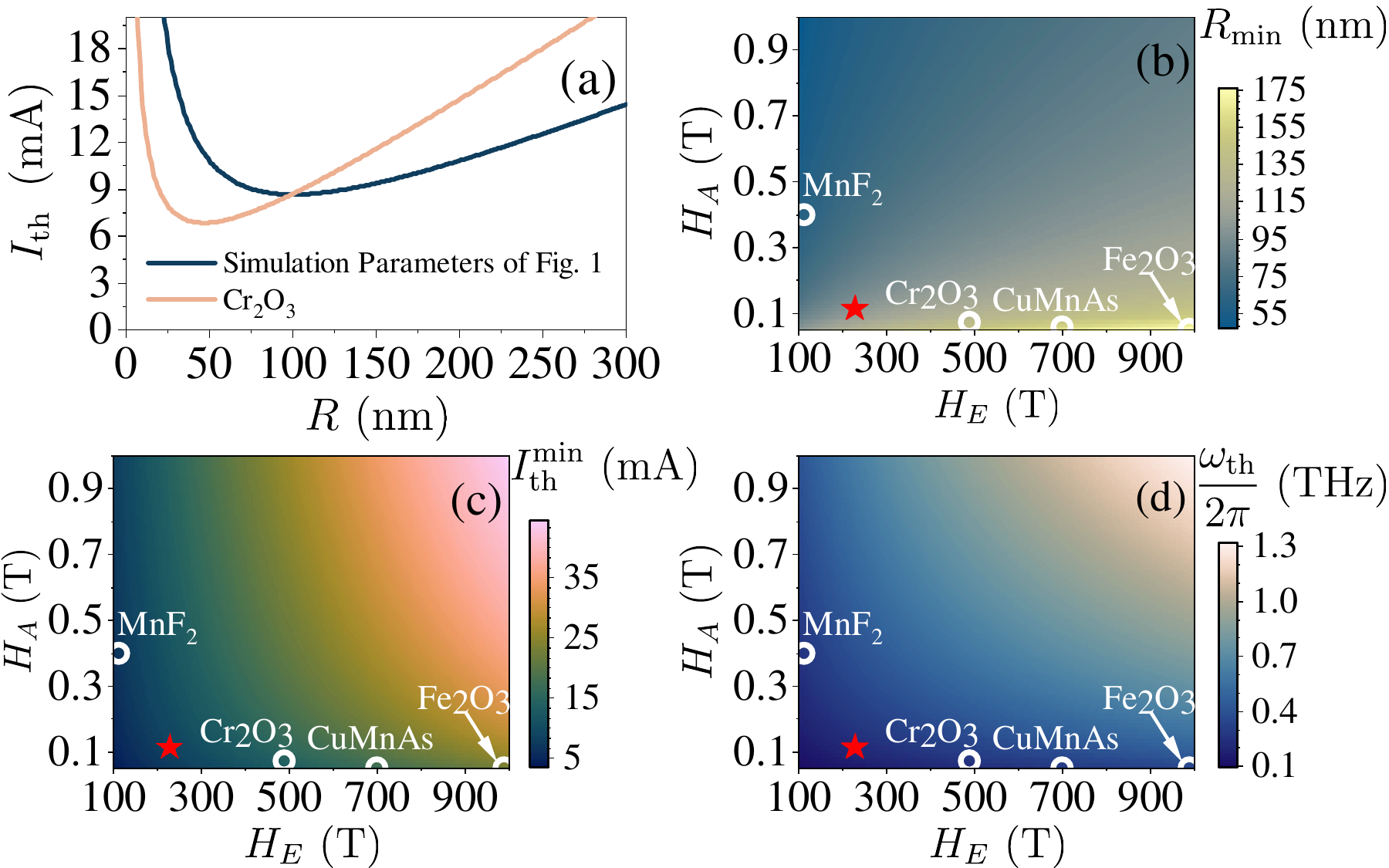}
	\caption{(a) Threshold current, $I_{\text{th}}$, as a function of $R$ for the parameters used in Fig. \ref{figure1} (blue line). The orange line considers Cr$_2$O$_3$ material parameters with the same HM and interface parameters as Fig. \ref{figure1}. NC radius (b), threshold current (c) and frequency (d) at the minimum point of $I_{\text{th}}(R)$ as a function of exchange and anisotropy fields. For all graphs we considered $d_{AF}=20$ nm as well as the corresponding $\alpha_{\text{SP}}$. {\color{black} Red asterisks indicate the material parameters used in Fig. \ref{figure1}. White circles highlight further AFMs.}}
	\label{figure2}
\end{figure}
\newline
\indent For practical applications the power consumption of the SHNO is of crucial importance. Approximating the current passing part of the NC in HM as a resistance with the length $L=2R$ and cross-sectional area $S=2R d_N$ in the $xz-$plane we obtain its resistance as $\sim(L/\sigma_N S)=1/\sigma_N d_N$ which is independent of $R$. By multiplying Eq. \ref{ThresholdCurrentDensity} by $S$ we obtain the threshold current as
\begin{equation}\label{ThresholdCurrent}
I_{\text{th}}(R)=\alpha\mathcal{A}\omega_{\text{th}}R+(\mathcal{B}\lambda_{ex}^2/R).
\end{equation}
The first term in this equation represents the current needed to overcome the damping which grows linearly with $R$. The second term reflects the propagation energy which scales with $R^{-1}$. The combination of these two terms leads to a minimum threshold current, $I_{\text{th}}^{\min}$, with respect to $R$. In Fig. \ref{figure2}(a) we depict $I_{\text{th}}$ as a function of $R$ for the parameters used in Fig. \ref{figure1} (blue line). For comparison, we display $I_{\text{th}}(R)$ for the uniaxial antiferromagnet Cr$_2$O$_3$ (orange line) with material parameters of $M_s=45.5$ kA/m, $H_E=490$ T, $H_A=0.035$ T and $a=1$ nm \cite{Foner1963,Artman1965}. Since $\omega_{\text{th}}$ is a slowly varying function of $R$ for $R\gtrsim 2\lambda_{ex}$ (Fig. \ref{figure1}(d)) away from the exchange dominated regime, we neglect its dependency on $R$, for large $R$ and obtain  $R_{\text{min}}\simeq\sqrt{\frac{\mathcal{B}\lambda_{ex}^2}{\alpha\mathcal{A}\omega_{\text{th}}^{\text{min}}}}$ where $R_{\text{min}}$ and $\omega_{\text{th}}^{\text{min}}$ are the NC radius and threshold frequency at the threshold current minimum, respectively. This expression slightly underestimates $R_{\text{min}}$. In Fig. \ref{figure2}(b), (c) and (d) we depict $R_{\text{min}}$, $I_{\text{th}}^{\min}$ and $\omega_{\text{th}}^{\text{min}}$, respectively, as a function of $H_E$ and $H_A$ {\color{black} (for a set of parameters that hold the exchange limit)} which are numerically obtained by minimizing $I_{\text{th}}(R)$. Figure \ref{figure2} (b) shows that $R_{\text{min}}$ increases (decreases) by increasing $H_E$ ($H_A$). However, $I_{\text{th}}^{\min}$ (Fig. \ref{figure2} (c)) and $\omega_{\text{th}}^{\text{min}}$ (Fig. \ref{figure2} (d)) increase by increasing both $H_E$ and $H_A$. {\color{black} The parameters used for simulations is indicated with the red asterisk. White circles indicate some of the most interesting materials for AFM spintronics \cite{Foner1963,Artman1965,Saidl2017,Lebrun2018,Wang2020,Li2020,Vaidya2020,Fink1964,Searle1968,Moriyama2019,Lebrun2020,Hamdi2022a,Zhou2022} which all of them fall in the exchange limit and our theory holds.}
\newline
\indent Due to the large exchange field in AFMs, the effect of damping on $I_{\text{th}}(R)$ is enhanced (Eq. \ref{ThresholdCurrentDensity} and \ref{ThresholdCurrent}). Therefore, we evaluate the effect of AFM thickness which drastically influences $\alpha_{\text{SP}}$. The intrinsic Gilbert damping, $\alpha_0$, is in the order of $10^{-6}-10^{-4}$ for insulating AFMs \cite{Fink1964,Searle1968,Moriyama2019,Lebrun2020,Hamdi2022a} which is negligibly small compared to $\alpha_{\text{SP}}$ for thin films.
\begin{figure}[t]
	\centering
	\includegraphics[width=0.48\textwidth]{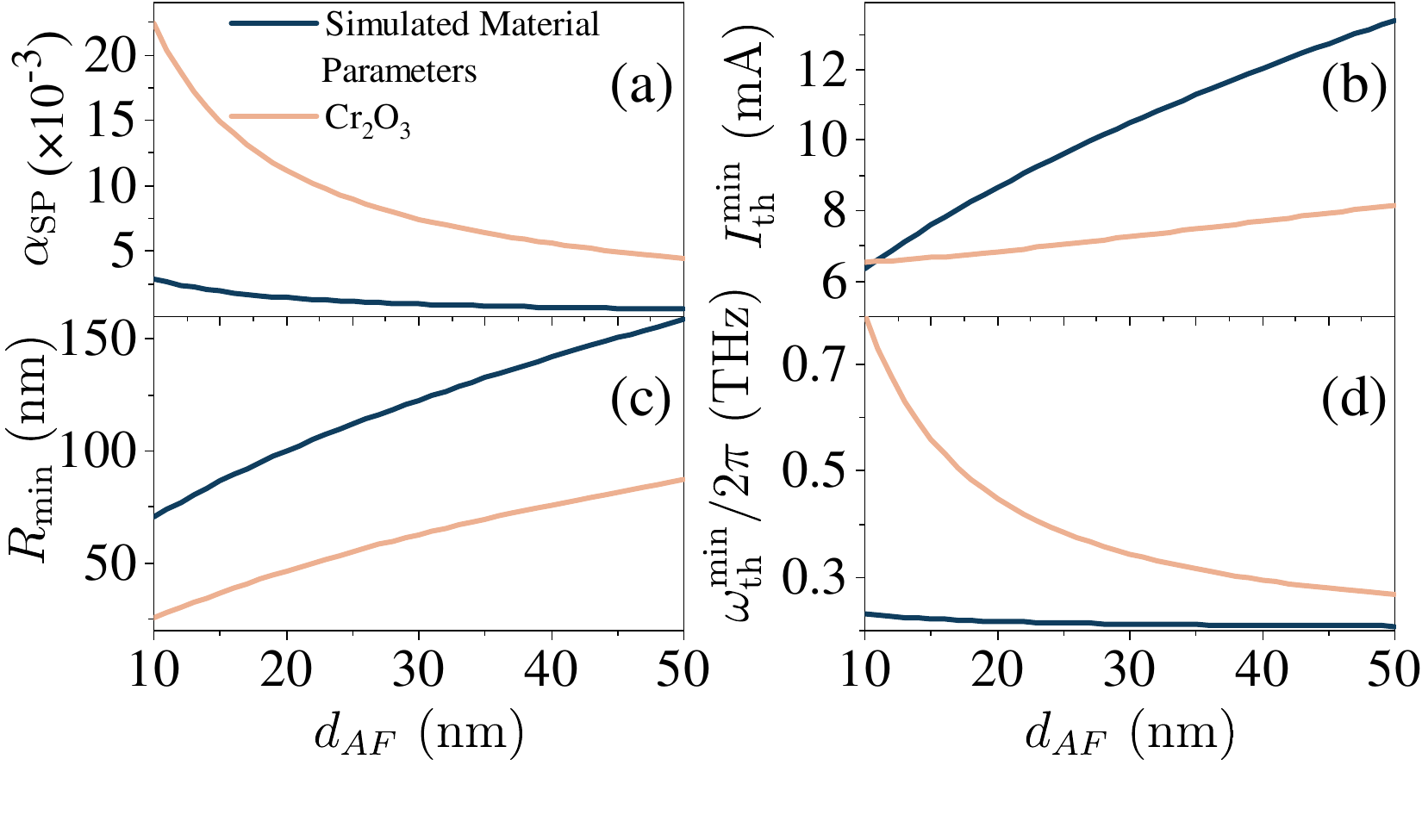}
	\caption{Spin pumping induced damping, $\alpha_{\text{SP}}$ (a) threshold current $I_{\text{th}}^{\min}$ (b), NC size $R_{\text{min}}$ (c) and threshold frequency, $\omega_{\text{th}}^{\text{min}}$ (d) at the threshold current minimum as a function of AFM layer thickness,  $d_{AF}$.}
	\label{figure3}
\end{figure}
Figure \ref{figure3}(a) shows $\alpha_{\text{SP}}$ for the AFM simulated in Fig. \ref{figure1} and Cr$_2$O$_3$. Since $M_s$ for Cr$_2$O$_3$ is small, $\alpha_{\text{SP}}$ depends prominently on $d_{AF}$. Despite the large $\alpha_{\text{SP}}$ for Cr$_2$O$_3$, its $I_{\text{th}}^{\min}$ (orange curve) is considerably smaller compared to the AFM with large $M_s$ (blue curve) as is shown in Fig. \ref{figure3}(b). Though $\alpha_{\text{SP}}$ decreases by increasing $d_{AF}$, $I_{\text{th}}^{\min}$ increases since SOT is an interface effect and scales with $d_{AF}^{-1}$. $R_{\text{min}}$ increases as a function of $d_{AF}$ (Fig. \ref{figure3}(c)) since $R_{\text{min}}\propto \sqrt{\alpha_{\text{SP}}^{-1}}\propto\sqrt{d_{AF}}$ and $\mathcal{A}$ and $\mathcal{B}$ are slowly varying functions of $\alpha$. Despite having larger $H_E$ and smaller $H_A$, Cr$_2$O$_3$ has smaller $R_{\text{min}}$ due to larger $\alpha_{\text{SP}}$ and $\omega_{\text{th}}^{\text{min}}$. Though the AFMR frequency of Cr$_2$O$_3$ (164 GHz) is lower than the AFMR of Fig. \ref{figure1} (202.5 GHz), the threshold frequency at $I_{\text{th}}^{\min}$ is higher for Cr$_2$O$_3$ (Fig. \ref{figure3}(d)). The higher frequency is a consequence of low $M_s$ which results in a significant change of $\alpha_{\text{SP}}$ with $d_{AF}$.\\
\indent {\color{black} In the following we make use of the predicted spin dynamics and evaluate electrical signals that arise due to spin pumping into an adjacent HM layer. The dynamic part of the N\'eel vector, $\mathbf{n}(r,t)$, gives rise to spin pumping which scales with $\mathbf{j}_{s}=\left(\mathbf{n}\times\mathbf{\dot{n}}\right)\times\hat{\mathbf{z}}$ \cite{Cheng2014,Cheng2016,Khymyn2017a}. By adding a HM layer either above or below the AFM, an ac electrical field is generated according to $\mathbf{E}=\mathcal{E}\mathbf{j}_{s}=\mathcal{E}\left(\mathbf{n}\times\dot{\mathbf{n}}\right)\times \mathbf{\hat{z}}$ with $\mathcal{E}=\hbar\eta \theta_{SH}/e d_N $ \cite{Cheng2016}. Taking the complex SW amplitude, $u$, we obtain the ac signal \cite{SupplementalNote1}
\begin{eqnarray}\label{SpinPumping}
E_{\rm{ac}}(r,t)&=&\mathcal{E}n_{x}\dot{n}_z=\mathcal{E}\omega\sqrt{1-|u(r)|^{2}}\,\text{Re}\left\{u(r) e^{i\omega t}\right\}. \label{Eac}
\end{eqnarray}
We numerically calculated $\mathbf{j}_{s}=\left(\mathbf{n}\times\mathbf{\dot{n}}\right)\times\hat{\mathbf{z}}$ for each cell at each moment in time from micromagnetic simulations. We took the FFT of this quantity, multiplied by $\mathcal{E}=0.7\times 10^{-7}$ V.s/cm and color-coded the absolute values as shown for two different currents $J_c$ applied to a NC with $R=25$ nm in Fig. \ref{figure4}(a) and (b). Thereby, we evaluate the spin pumping at $f_{\text{th}}=0.3$ THz and $f_{\text{h}}=1.2$ THz, respectively. We display its spatially inhomogeneous characteristics in units of V/cm. The panels of Fig. \ref{figure4}(a) and (b) hence show ac electric fields in the typical HM assumed above. They allow for a straightforward comparison with the earlier reports on spin pumping from biaxial AFMs. For the $f_{\text{th}}$ mode, the maximum spin pumping signal is obtained in the NC, while for the $f_{\text{h}}$ mode, $E_{\text{ac}}$ vanishes under the NC. This result is explained by Fig. \ref{figure4}(c) illustrating the spin-precessional motion at $f_{\text{th}}$ (blue) and $f_{\text{h}}$ (red) in the center of the NC. For $f_{\text{th}}$, the SOT excites a conically precessing N\'eel vector (See Supplemental video 1 \cite{SupplementalNote1}) with a well-defined almost circular polarization and a cone angle smaller than $\pi/2$. At large $J_c$ (SOT), the cone angle reaches $\pi/2$ and the N\'eel vector precesses entirely in the $yz$-plane (See Supplemental video 2 \cite{SupplementalNote1}). In this case, the spin-precessional amplitude $|u(r)|$ is maximum ($|u(r)|=1$). In Fig. \ref{figure4}(d) we depict the quantity $j_{s,ac}=|u(r)|\sqrt{1-|u(r)|^{2}}$ as a function of $|u(r)|$. The striking feature of $j_{s,ac}$ is the maximum ac spin pumping at $|u|_{\text{max}}=\sqrt{2}/2$. Consistent with the macrospin approximation applied in Ref. \onlinecite{Khymyn2017a}, we observe zero spin pumping inside the NC when a large $J_c$ gives rise to $|u(r)|=1$. But, the zero-spin pumping is no longer true for the discovered Slonzweski waves propagating into the AFM. Outside the NC a region is found where due to damping the amplitude $|u|_{\text{max}}=\sqrt{2}/2$ is realized and maximum ac spin pumping occurs remotely from the NC. The predicted excitation of propagating THz-ASSWs makes uniaxial AFMs a promising material for SHNOs as they exhibit a lower threshold current and enable 3 orders of magnitude larger values $|E_{ac}|$ than the earlier considered biaxial AFMs \cite{Cheng2016,Khymyn2017a}. In Eq. \ref{Eac}, we find $|E_{ac}|$ to be proportional to the operational frequency $\omega$, whereas for biaxial AFMs one finds $|E_{ac}|\propto \frac{\omega_{e}\omega_{E}}{4\sqrt{\alpha^{2}\omega_{E}^{2}+\omega^2}}$. The latter value is orders of magnitude smaller due to small biaxial anisotropy $\omega_{e}$.
\begin{figure}[t]
	\centering
	\includegraphics[width=0.45\textwidth]{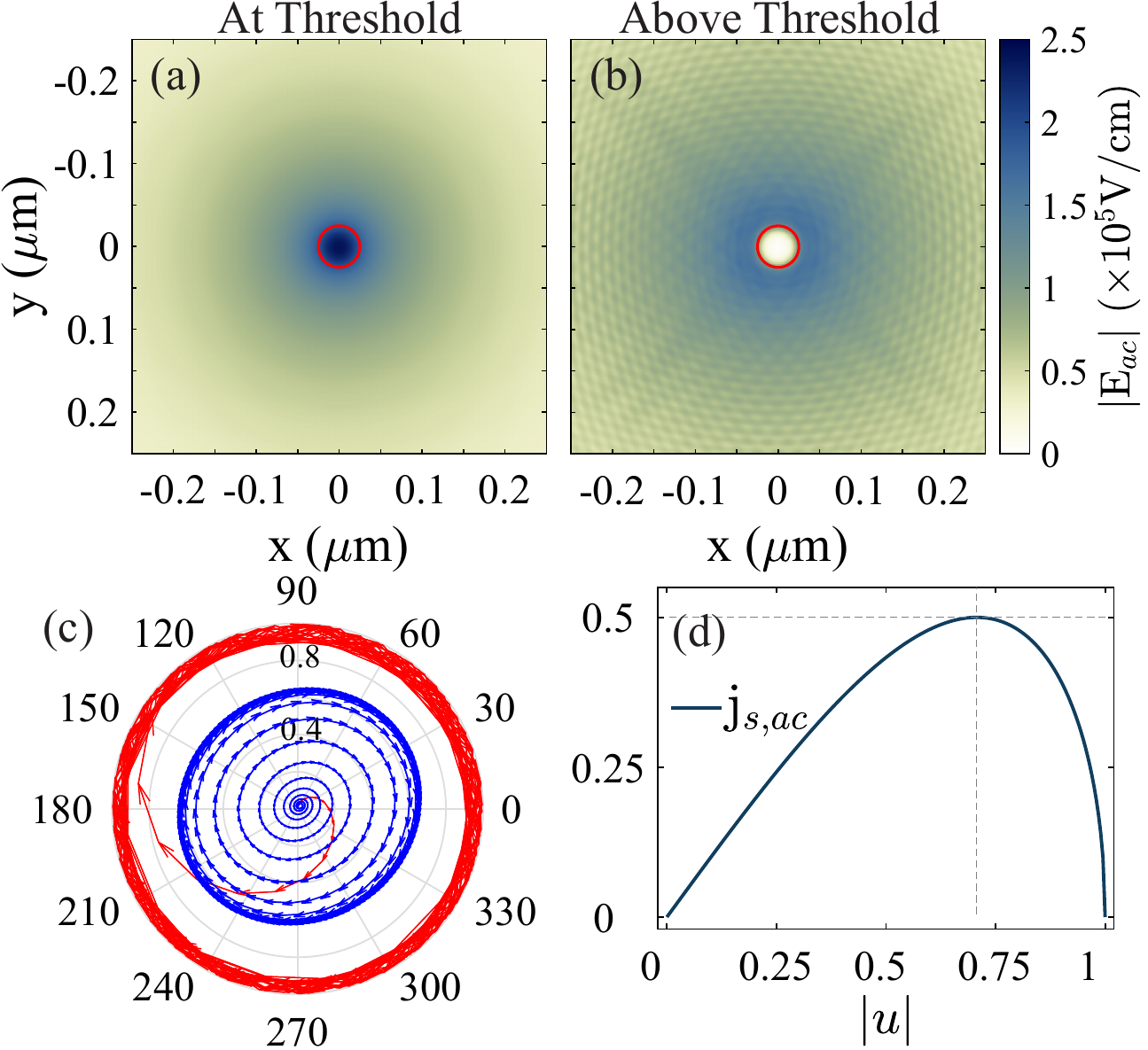}
	\caption{{\color{black}Spin pumping induced ac output electric field for modes (a) $f_{\text{th}}$ at threshold and (b) $f_{\text{h}}$ above threshold. Red circule marks the NC. (c) Time dependent dynamic N\'eel vector for $f_{\text{th}}$ (blue) and $f_{\text{h}}$ (red). The arrows indicate the direction of time. (d) The dimensionless parameter $j_{s,ac}=|u|\sqrt{1-|u|^{2}}$ as a function of the spin wave amplitude $|u|$.}}
	\label{figure4}
\end{figure}}\\ 
\indent \textit{Conclusion.}--- We have derived and solved the spin wave equation for a realistic SHNO considering AFMs with uni-axial anisotropy. We find propagating THz Slonzewski waves which allow one to generate large ac electrical fields via spin pumping into an adjacent heavy metal film. At the same time, synchronization between neighboring SHNOs via propagating SWs might arise. The predicted waves hence open unprecedented pathways for dc current driven THz signal generation by means of magnons in thin-film AFMs. \\
\indent \textit{Acknowledgment.}--- We thank SNSF for financial support via grant 177550.
\bibliographystyle{apsrev4-2}

\begin{thebibliography}{76}%
	\makeatletter
	\providecommand \@ifxundefined [1]{%
		\@ifx{#1\undefined}
	}%
	\providecommand \@ifnum [1]{%
		\ifnum #1\expandafter \@firstoftwo
		\else \expandafter \@secondoftwo
		\fi
	}%
	\providecommand \@ifx [1]{%
		\ifx #1\expandafter \@firstoftwo
		\else \expandafter \@secondoftwo
		\fi
	}%
	\providecommand \natexlab [1]{#1}%
	\providecommand \enquote  [1]{``#1''}%
	\providecommand \bibnamefont  [1]{#1}%
	\providecommand \bibfnamefont [1]{#1}%
	\providecommand \citenamefont [1]{#1}%
	\providecommand \href@noop [0]{\@secondoftwo}%
	\providecommand \href [0]{\begingroup \@sanitize@url \@href}%
	\providecommand \@href[1]{\@@startlink{#1}\@@href}%
	\providecommand \@@href[1]{\endgroup#1\@@endlink}%
	\providecommand \@sanitize@url [0]{\catcode `\\12\catcode `\$12\catcode
		`\&12\catcode `\#12\catcode `\^12\catcode `\_12\catcode `\%12\relax}%
	\providecommand \@@startlink[1]{}%
	\providecommand \@@endlink[0]{}%
	\providecommand \url  [0]{\begingroup\@sanitize@url \@url }%
	\providecommand \@url [1]{\endgroup\@href {#1}{\urlprefix }}%
	\providecommand \urlprefix  [0]{URL }%
	\providecommand \Eprint [0]{\href }%
	\providecommand \doibase [0]{https://doi.org/}%
	\providecommand \selectlanguage [0]{\@gobble}%
	\providecommand \bibinfo  [0]{\@secondoftwo}%
	\providecommand \bibfield  [0]{\@secondoftwo}%
	\providecommand \translation [1]{[#1]}%
	\providecommand \BibitemOpen [0]{}%
	\providecommand \bibitemStop [0]{}%
	\providecommand \bibitemNoStop [0]{.\EOS\space}%
	\providecommand \EOS [0]{\spacefactor3000\relax}%
	\providecommand \BibitemShut  [1]{\csname bibitem#1\endcsname}%
	\let\auto@bib@innerbib\@empty
	\bibitem [{\citenamefont {Nagatsuma}\ \emph {et~al.}(2016)\citenamefont
		{Nagatsuma}, \citenamefont {Ducournau},\ and\ \citenamefont
		{Renaud}}]{Nagatsuma2016}%
	\BibitemOpen
	\bibfield  {author} {\bibinfo {author} {\bibfnamefont {T.}~\bibnamefont
			{Nagatsuma}}, \bibinfo {author} {\bibfnamefont {G.}~\bibnamefont
			{Ducournau}},\ and\ \bibinfo {author} {\bibfnamefont {C.~C.}\ \bibnamefont
			{Renaud}},\ }\href {https://doi.org/10.1038/nphoton.2016.65} {\bibfield
		{journal} {\bibinfo  {journal} {Nat. Photonics}\ }\textbf {\bibinfo {volume}
			{10}},\ \bibinfo {pages} {371} (\bibinfo {year} {2016})}\BibitemShut
	{NoStop}%
	\bibitem [{\citenamefont {Ma}\ \emph {et~al.}(2019)\citenamefont {Ma},
		\citenamefont {Geng}, \citenamefont {Fan}, \citenamefont {Liu},\ and\
		\citenamefont {Chen}}]{Ma2019}%
	\BibitemOpen
	\bibfield  {author} {\bibinfo {author} {\bibfnamefont {Z.~T.}\ \bibnamefont
			{Ma}}, \bibinfo {author} {\bibfnamefont {Z.~X.}\ \bibnamefont {Geng}},
		\bibinfo {author} {\bibfnamefont {Z.~Y.}\ \bibnamefont {Fan}}, \bibinfo
		{author} {\bibfnamefont {J.}~\bibnamefont {Liu}},\ and\ \bibinfo {author}
		{\bibfnamefont {H.~D.}\ \bibnamefont {Chen}},\ }\href
	{https://doi.org/10.34133/2019/6482975} {\bibfield  {journal} {\bibinfo
			{journal} {Research}\ }\textbf {\bibinfo {volume} {2019}},\ \bibinfo {pages}
		{1} (\bibinfo {year} {2019})}\BibitemShut {NoStop}%
	\bibitem [{\citenamefont {Sirtori}(2002)}]{Sirtori2002}%
	\BibitemOpen
	\bibfield  {author} {\bibinfo {author} {\bibfnamefont {C.}~\bibnamefont
			{Sirtori}},\ }\href {https://doi.org/10.1038/417132b} {\bibfield  {journal}
		{\bibinfo  {journal} {Nature}\ }\textbf {\bibinfo {volume} {417}},\ \bibinfo
		{pages} {132} (\bibinfo {year} {2002})}\BibitemShut {NoStop}%
	\bibitem [{\citenamefont {Osborne}(2008)}]{Osborne2008}%
	\BibitemOpen
	\bibfield  {author} {\bibinfo {author} {\bibfnamefont {I.~S.}\ \bibnamefont
			{Osborne}},\ }\href {https://doi.org/10.1126/science.320.5881.1262b}
	{\bibfield  {journal} {\bibinfo  {journal} {Science (80-. ).}\ }\textbf
		{\bibinfo {volume} {320}},\ \bibinfo {pages} {1262b} (\bibinfo {year}
		{2008})}\BibitemShut {NoStop}%
	\bibitem [{\citenamefont {Cheng}\ \emph {et~al.}(2014)\citenamefont {Cheng},
		\citenamefont {Xiao}, \citenamefont {Niu},\ and\ \citenamefont
		{Brataas}}]{Cheng2014}%
	\BibitemOpen
	\bibfield  {author} {\bibinfo {author} {\bibfnamefont {R.}~\bibnamefont
			{Cheng}}, \bibinfo {author} {\bibfnamefont {J.}~\bibnamefont {Xiao}},
		\bibinfo {author} {\bibfnamefont {Q.}~\bibnamefont {Niu}},\ and\ \bibinfo
		{author} {\bibfnamefont {A.}~\bibnamefont {Brataas}},\ }\href
	{https://doi.org/10.1103/PhysRevLett.113.057601} {\bibfield  {journal}
		{\bibinfo  {journal} {Phys. Rev. Lett.}\ }\textbf {\bibinfo {volume} {113}},\
		\bibinfo {pages} {057601} (\bibinfo {year} {2014})}\BibitemShut {NoStop}%
	\bibitem [{\citenamefont {Jungwirth}\ \emph {et~al.}(2016)\citenamefont
		{Jungwirth}, \citenamefont {Marti}, \citenamefont {Wadley},\ and\
		\citenamefont {Wunderlich}}]{Jungwirth2016}%
	\BibitemOpen
	\bibfield  {author} {\bibinfo {author} {\bibfnamefont {T.}~\bibnamefont
			{Jungwirth}}, \bibinfo {author} {\bibfnamefont {X.}~\bibnamefont {Marti}},
		\bibinfo {author} {\bibfnamefont {P.}~\bibnamefont {Wadley}},\ and\ \bibinfo
		{author} {\bibfnamefont {J.}~\bibnamefont {Wunderlich}},\ }\href
	{https://doi.org/10.1038/nnano.2016.18} {\bibfield  {journal} {\bibinfo
			{journal} {Nat. Nanotechnol.}\ }\textbf {\bibinfo {volume} {11}},\ \bibinfo
		{pages} {231} (\bibinfo {year} {2016})}\BibitemShut {NoStop}%
	\bibitem [{\citenamefont {Baltz}\ \emph {et~al.}(2018)\citenamefont {Baltz},
		\citenamefont {Manchon}, \citenamefont {Tsoi}, \citenamefont {Moriyama},
		\citenamefont {Ono},\ and\ \citenamefont {Tserkovnyak}}]{Baltz2018}%
	\BibitemOpen
	\bibfield  {author} {\bibinfo {author} {\bibfnamefont {V.}~\bibnamefont
			{Baltz}}, \bibinfo {author} {\bibfnamefont {A.}~\bibnamefont {Manchon}},
		\bibinfo {author} {\bibfnamefont {M.}~\bibnamefont {Tsoi}}, \bibinfo {author}
		{\bibfnamefont {T.}~\bibnamefont {Moriyama}}, \bibinfo {author}
		{\bibfnamefont {T.}~\bibnamefont {Ono}},\ and\ \bibinfo {author}
		{\bibfnamefont {Y.}~\bibnamefont {Tserkovnyak}},\ }\href
	{https://doi.org/10.1103/RevModPhys.90.015005} {\bibfield  {journal}
		{\bibinfo  {journal} {Rev. Mod. Phys.}\ }\textbf {\bibinfo {volume} {90}},\
		\bibinfo {pages} {015005} (\bibinfo {year} {2018})}\BibitemShut {NoStop}%
	\bibitem [{\citenamefont {Jungwirth}\ \emph {et~al.}(2018)\citenamefont
		{Jungwirth}, \citenamefont {Sinova}, \citenamefont {Manchon}, \citenamefont
		{Marti}, \citenamefont {Wunderlich},\ and\ \citenamefont
		{Felser}}]{Jungwirth2018}%
	\BibitemOpen
	\bibfield  {author} {\bibinfo {author} {\bibfnamefont {T.}~\bibnamefont
			{Jungwirth}}, \bibinfo {author} {\bibfnamefont {J.}~\bibnamefont {Sinova}},
		\bibinfo {author} {\bibfnamefont {A.}~\bibnamefont {Manchon}}, \bibinfo
		{author} {\bibfnamefont {X.}~\bibnamefont {Marti}}, \bibinfo {author}
		{\bibfnamefont {J.}~\bibnamefont {Wunderlich}},\ and\ \bibinfo {author}
		{\bibfnamefont {C.}~\bibnamefont {Felser}},\ }\href
	{https://doi.org/10.1038/s41567-018-0063-6} {\bibfield  {journal} {\bibinfo
			{journal} {Nat. Phys. 2018 143}\ }\textbf {\bibinfo {volume} {14}},\ \bibinfo
		{pages} {200} (\bibinfo {year} {2018})}\BibitemShut {NoStop}%
	\bibitem [{\citenamefont {Duine}\ \emph {et~al.}(2018)\citenamefont {Duine},
		\citenamefont {Lee}, \citenamefont {Parkin},\ and\ \citenamefont
		{Stiles}}]{Duine2018}%
	\BibitemOpen
	\bibfield  {author} {\bibinfo {author} {\bibfnamefont {R.~A.}\ \bibnamefont
			{Duine}}, \bibinfo {author} {\bibfnamefont {K.~J.}\ \bibnamefont {Lee}},
		\bibinfo {author} {\bibfnamefont {S.~S.}\ \bibnamefont {Parkin}},\ and\
		\bibinfo {author} {\bibfnamefont {M.~D.}\ \bibnamefont {Stiles}},\ }\href
	{https://doi.org/10.1038/s41567-018-0050-y} {\bibfield  {journal} {\bibinfo
			{journal} {Nat. Phys. 2018 143}\ }\textbf {\bibinfo {volume} {14}},\ \bibinfo
		{pages} {217} (\bibinfo {year} {2018})}\BibitemShut {NoStop}%
	\bibitem [{\citenamefont {Gomonay}\ \emph
		{et~al.}(2018{\natexlab{a}})\citenamefont {Gomonay}, \citenamefont {Baltz},
		\citenamefont {Brataas},\ and\ \citenamefont {Tserkovnyak}}]{Gomonay2018a}%
	\BibitemOpen
	\bibfield  {author} {\bibinfo {author} {\bibfnamefont {O.}~\bibnamefont
			{Gomonay}}, \bibinfo {author} {\bibfnamefont {V.}~\bibnamefont {Baltz}},
		\bibinfo {author} {\bibfnamefont {A.}~\bibnamefont {Brataas}},\ and\ \bibinfo
		{author} {\bibfnamefont {Y.}~\bibnamefont {Tserkovnyak}},\ }\href
	{https://doi.org/10.1038/s41567-018-0049-4} {\bibfield  {journal} {\bibinfo
			{journal} {Nat. Phys. 2018 143}\ }\textbf {\bibinfo {volume} {14}},\ \bibinfo
		{pages} {213} (\bibinfo {year} {2018}{\natexlab{a}})}\BibitemShut {NoStop}%
	\bibitem [{\citenamefont {{\v{Z}}elezn{\'{y}}}\ \emph
		{et~al.}(2018)\citenamefont {{\v{Z}}elezn{\'{y}}}, \citenamefont {Wadley},
		\citenamefont {Olejn{\'{i}}k}, \citenamefont {Hoffmann},\ and\ \citenamefont
		{Ohno}}]{Zelezny2018}%
	\BibitemOpen
	\bibfield  {author} {\bibinfo {author} {\bibfnamefont {J.}~\bibnamefont
			{{\v{Z}}elezn{\'{y}}}}, \bibinfo {author} {\bibfnamefont {P.}~\bibnamefont
			{Wadley}}, \bibinfo {author} {\bibfnamefont {K.}~\bibnamefont
			{Olejn{\'{i}}k}}, \bibinfo {author} {\bibfnamefont {A.}~\bibnamefont
			{Hoffmann}},\ and\ \bibinfo {author} {\bibfnamefont {H.}~\bibnamefont
			{Ohno}},\ }\href {https://doi.org/10.1038/s41567-018-0062-7} {\bibfield
		{journal} {\bibinfo  {journal} {Nat. Phys. 2018 143}\ }\textbf {\bibinfo
			{volume} {14}},\ \bibinfo {pages} {220} (\bibinfo {year} {2018})}\BibitemShut
	{NoStop}%
	\bibitem [{\citenamefont {N{\v{e}}mec}\ \emph {et~al.}(2018)\citenamefont
		{N{\v{e}}mec}, \citenamefont {Fiebig}, \citenamefont {Kampfrath},\ and\
		\citenamefont {Kimel}}]{Nemec2018}%
	\BibitemOpen
	\bibfield  {author} {\bibinfo {author} {\bibfnamefont {P.}~\bibnamefont
			{N{\v{e}}mec}}, \bibinfo {author} {\bibfnamefont {M.}~\bibnamefont {Fiebig}},
		\bibinfo {author} {\bibfnamefont {T.}~\bibnamefont {Kampfrath}},\ and\
		\bibinfo {author} {\bibfnamefont {A.~V.}\ \bibnamefont {Kimel}},\ }\href
	{https://doi.org/10.1038/s41567-018-0051-x} {\bibfield  {journal} {\bibinfo
			{journal} {Nat. Phys. 2018 143}\ }\textbf {\bibinfo {volume} {14}},\ \bibinfo
		{pages} {229} (\bibinfo {year} {2018})}\BibitemShut {NoStop}%
	\bibitem [{\citenamefont {{\v{S}}mejkal}\ \emph {et~al.}(2018)\citenamefont
		{{\v{S}}mejkal}, \citenamefont {Mokrousov}, \citenamefont {Yan},\ and\
		\citenamefont {MacDonald}}]{Smejkal2018}%
	\BibitemOpen
	\bibfield  {author} {\bibinfo {author} {\bibfnamefont {L.}~\bibnamefont
			{{\v{S}}mejkal}}, \bibinfo {author} {\bibfnamefont {Y.}~\bibnamefont
			{Mokrousov}}, \bibinfo {author} {\bibfnamefont {B.}~\bibnamefont {Yan}},\
		and\ \bibinfo {author} {\bibfnamefont {A.~H.}\ \bibnamefont {MacDonald}},\
	}\href {https://doi.org/10.1038/s41567-018-0064-5} {\bibfield  {journal}
		{\bibinfo  {journal} {Nat. Phys. 2018 143}\ }\textbf {\bibinfo {volume}
			{14}},\ \bibinfo {pages} {242} (\bibinfo {year} {2018})}\BibitemShut
	{NoStop}%
	\bibitem [{\citenamefont {Cheng}\ \emph {et~al.}(2016)\citenamefont {Cheng},
		\citenamefont {Xiao},\ and\ \citenamefont {Brataas}}]{Cheng2016}%
	\BibitemOpen
	\bibfield  {author} {\bibinfo {author} {\bibfnamefont {R.}~\bibnamefont
			{Cheng}}, \bibinfo {author} {\bibfnamefont {D.}~\bibnamefont {Xiao}},\ and\
		\bibinfo {author} {\bibfnamefont {A.}~\bibnamefont {Brataas}},\ }\href
	{https://doi.org/10.1103/PhysRevLett.116.207603} {\bibfield  {journal}
		{\bibinfo  {journal} {Phys. Rev. Lett.}\ }\textbf {\bibinfo {volume} {116}},\
		\bibinfo {pages} {207603} (\bibinfo {year} {2016})}\BibitemShut {NoStop}%
	\bibitem [{\citenamefont {Khymyn}\ \emph
		{et~al.}(2017{\natexlab{a}})\citenamefont {Khymyn}, \citenamefont {Lisenkov},
		\citenamefont {Tiberkevich}, \citenamefont {Ivanov},\ and\ \citenamefont
		{Slavin}}]{Khymyn2017a}%
	\BibitemOpen
	\bibfield  {author} {\bibinfo {author} {\bibfnamefont {R.}~\bibnamefont
			{Khymyn}}, \bibinfo {author} {\bibfnamefont {I.}~\bibnamefont {Lisenkov}},
		\bibinfo {author} {\bibfnamefont {V.}~\bibnamefont {Tiberkevich}}, \bibinfo
		{author} {\bibfnamefont {B.~A.}\ \bibnamefont {Ivanov}},\ and\ \bibinfo
		{author} {\bibfnamefont {A.}~\bibnamefont {Slavin}},\ }\href
	{https://doi.org/10.1038/srep43705} {\bibfield  {journal} {\bibinfo
			{journal} {Sci. Rep.}\ }\textbf {\bibinfo {volume} {7}},\ \bibinfo {pages}
		{43705} (\bibinfo {year} {2017}{\natexlab{a}})}\BibitemShut {NoStop}%
	\bibitem [{\citenamefont {Sulymenko}\ \emph {et~al.}(2017)\citenamefont
		{Sulymenko}, \citenamefont {Prokopenko}, \citenamefont {Tiberkevich},
		\citenamefont {Slavin}, \citenamefont {Ivanov},\ and\ \citenamefont
		{Khymyn}}]{Sulymenko2017}%
	\BibitemOpen
	\bibfield  {author} {\bibinfo {author} {\bibfnamefont {O.~R.}\ \bibnamefont
			{Sulymenko}}, \bibinfo {author} {\bibfnamefont {O.~V.}\ \bibnamefont
			{Prokopenko}}, \bibinfo {author} {\bibfnamefont {V.~S.}\ \bibnamefont
			{Tiberkevich}}, \bibinfo {author} {\bibfnamefont {A.~N.}\ \bibnamefont
			{Slavin}}, \bibinfo {author} {\bibfnamefont {B.~A.}\ \bibnamefont {Ivanov}},\
		and\ \bibinfo {author} {\bibfnamefont {R.~S.}\ \bibnamefont {Khymyn}},\
	}\href {https://doi.org/10.1103/PhysRevApplied.8.064007} {\bibfield
		{journal} {\bibinfo  {journal} {Phys. Rev. Appl.}\ }\textbf {\bibinfo
			{volume} {8}},\ \bibinfo {pages} {064007} (\bibinfo {year}
		{2017})}\BibitemShut {NoStop}%
	\bibitem [{\citenamefont {Johansen}\ and\ \citenamefont
		{Brataas}(2017)}]{Johansen2017}%
	\BibitemOpen
	\bibfield  {author} {\bibinfo {author} {\bibfnamefont {{\O}.}~\bibnamefont
			{Johansen}}\ and\ \bibinfo {author} {\bibfnamefont {A.}~\bibnamefont
			{Brataas}},\ }\href {https://doi.org/10.1103/PhysRevB.95.220408} {\bibfield
		{journal} {\bibinfo  {journal} {Phys. Rev. B}\ }\textbf {\bibinfo {volume}
			{95}},\ \bibinfo {pages} {220408} (\bibinfo {year} {2017})}\BibitemShut
	{NoStop}%
	\bibitem [{\citenamefont {Ch{\c{e}}ci{\'{n}}ski}\ \emph
		{et~al.}(2017)\citenamefont {Ch{\c{e}}ci{\'{n}}ski}, \citenamefont
		{Frankowski},\ and\ \citenamefont {Stobiecki}}]{Checinski2017}%
	\BibitemOpen
	\bibfield  {author} {\bibinfo {author} {\bibfnamefont {J.}~\bibnamefont
			{Ch{\c{e}}ci{\'{n}}ski}}, \bibinfo {author} {\bibfnamefont {M.}~\bibnamefont
			{Frankowski}},\ and\ \bibinfo {author} {\bibfnamefont {T.}~\bibnamefont
			{Stobiecki}},\ }\href {https://doi.org/10.1103/PhysRevB.96.174438} {\bibfield
		{journal} {\bibinfo  {journal} {Phys. Rev. B}\ }\textbf {\bibinfo {volume}
			{96}},\ \bibinfo {pages} {174438} (\bibinfo {year} {2017})}\BibitemShut
	{NoStop}%
	\bibitem [{\citenamefont {Khymyn}\ \emph
		{et~al.}(2017{\natexlab{b}})\citenamefont {Khymyn}, \citenamefont
		{Tiberkevich},\ and\ \citenamefont {Slavin}}]{Khymyn2017}%
	\BibitemOpen
	\bibfield  {author} {\bibinfo {author} {\bibfnamefont {R.}~\bibnamefont
			{Khymyn}}, \bibinfo {author} {\bibfnamefont {V.}~\bibnamefont
			{Tiberkevich}},\ and\ \bibinfo {author} {\bibfnamefont {A.}~\bibnamefont
			{Slavin}},\ }\href {https://doi.org/10.1063/1.4977974} {\bibfield  {journal}
		{\bibinfo  {journal} {AIP Adv.}\ }\textbf {\bibinfo {volume} {7}},\ \bibinfo
		{pages} {055931} (\bibinfo {year} {2017}{\natexlab{b}})}\BibitemShut
	{NoStop}%
	\bibitem [{\citenamefont {Johansen}\ \emph {et~al.}(2018)\citenamefont
		{Johansen}, \citenamefont {Skarsv{\aa}g},\ and\ \citenamefont
		{Brataas}}]{Johansen2018}%
	\BibitemOpen
	\bibfield  {author} {\bibinfo {author} {\bibfnamefont {{\O}.}~\bibnamefont
			{Johansen}}, \bibinfo {author} {\bibfnamefont {H.}~\bibnamefont
			{Skarsv{\aa}g}},\ and\ \bibinfo {author} {\bibfnamefont {A.}~\bibnamefont
			{Brataas}},\ }\href {https://doi.org/10.1103/PhysRevB.97.054423} {\bibfield
		{journal} {\bibinfo  {journal} {Phys. Rev. B}\ }\textbf {\bibinfo {volume}
			{97}},\ \bibinfo {pages} {054423} (\bibinfo {year} {2018})}\BibitemShut
	{NoStop}%
	\bibitem [{\citenamefont {Gomonay}\ \emph
		{et~al.}(2018{\natexlab{b}})\citenamefont {Gomonay}, \citenamefont
		{Jungwirth},\ and\ \citenamefont {Sinova}}]{Gomonay2018}%
	\BibitemOpen
	\bibfield  {author} {\bibinfo {author} {\bibfnamefont {O.}~\bibnamefont
			{Gomonay}}, \bibinfo {author} {\bibfnamefont {T.}~\bibnamefont {Jungwirth}},\
		and\ \bibinfo {author} {\bibfnamefont {J.}~\bibnamefont {Sinova}},\ }\href
	{https://doi.org/10.1103/PhysRevB.98.104430} {\bibfield  {journal} {\bibinfo
			{journal} {Phys. Rev. B}\ }\textbf {\bibinfo {volume} {98}},\ \bibinfo
		{pages} {104430} (\bibinfo {year} {2018}{\natexlab{b}})}\BibitemShut
	{NoStop}%
	\bibitem [{\citenamefont {Puliafito}\ \emph {et~al.}(2019)\citenamefont
		{Puliafito}, \citenamefont {Khymyn}, \citenamefont {Carpentieri},
		\citenamefont {Azzerboni}, \citenamefont {Tiberkevich}, \citenamefont
		{Slavin},\ and\ \citenamefont {Finocchio}}]{Puliafito2019}%
	\BibitemOpen
	\bibfield  {author} {\bibinfo {author} {\bibfnamefont {V.}~\bibnamefont
			{Puliafito}}, \bibinfo {author} {\bibfnamefont {R.}~\bibnamefont {Khymyn}},
		\bibinfo {author} {\bibfnamefont {M.}~\bibnamefont {Carpentieri}}, \bibinfo
		{author} {\bibfnamefont {B.}~\bibnamefont {Azzerboni}}, \bibinfo {author}
		{\bibfnamefont {V.}~\bibnamefont {Tiberkevich}}, \bibinfo {author}
		{\bibfnamefont {A.}~\bibnamefont {Slavin}},\ and\ \bibinfo {author}
		{\bibfnamefont {G.}~\bibnamefont {Finocchio}},\ }\href
	{https://doi.org/10.1103/PhysRevB.99.024405} {\bibfield  {journal} {\bibinfo
			{journal} {Phys. Rev. B}\ }\textbf {\bibinfo {volume} {99}},\ \bibinfo
		{pages} {024405} (\bibinfo {year} {2019})}\BibitemShut {NoStop}%
	\bibitem [{\citenamefont {Lee}\ \emph {et~al.}(2019)\citenamefont {Lee},
		\citenamefont {Park},\ and\ \citenamefont {Lee}}]{Lee2019}%
	\BibitemOpen
	\bibfield  {author} {\bibinfo {author} {\bibfnamefont {D.-K.}\ \bibnamefont
			{Lee}}, \bibinfo {author} {\bibfnamefont {B.-G.}\ \bibnamefont {Park}},\ and\
		\bibinfo {author} {\bibfnamefont {K.-J.}\ \bibnamefont {Lee}},\ }\href
	{https://doi.org/10.1103/PhysRevApplied.11.054048} {\bibfield  {journal}
		{\bibinfo  {journal} {Phys. Rev. Appl.}\ }\textbf {\bibinfo {volume} {11}},\
		\bibinfo {pages} {054048} (\bibinfo {year} {2019})}\BibitemShut {NoStop}%
	\bibitem [{\citenamefont {Lisenkov}\ \emph {et~al.}(2019)\citenamefont
		{Lisenkov}, \citenamefont {Khymyn}, \citenamefont {{\AA}kerman},
		\citenamefont {Sun},\ and\ \citenamefont {Ivanov}}]{Lisenkov2019}%
	\BibitemOpen
	\bibfield  {author} {\bibinfo {author} {\bibfnamefont {I.}~\bibnamefont
			{Lisenkov}}, \bibinfo {author} {\bibfnamefont {R.}~\bibnamefont {Khymyn}},
		\bibinfo {author} {\bibfnamefont {J.}~\bibnamefont {{\AA}kerman}}, \bibinfo
		{author} {\bibfnamefont {N.~X.}\ \bibnamefont {Sun}},\ and\ \bibinfo {author}
		{\bibfnamefont {B.~A.}\ \bibnamefont {Ivanov}},\ }\href
	{https://doi.org/10.1103/PhysRevB.100.100409} {\bibfield  {journal} {\bibinfo
			{journal} {Phys. Rev. B}\ }\textbf {\bibinfo {volume} {100}},\ \bibinfo
		{pages} {100409} (\bibinfo {year} {2019})}\BibitemShut {NoStop}%
	\bibitem [{\citenamefont {Troncoso}\ \emph {et~al.}(2019)\citenamefont
		{Troncoso}, \citenamefont {Rode}, \citenamefont {Stamenov}, \citenamefont
		{Coey},\ and\ \citenamefont {Brataas}}]{Troncoso2019}%
	\BibitemOpen
	\bibfield  {author} {\bibinfo {author} {\bibfnamefont {R.~E.}\ \bibnamefont
			{Troncoso}}, \bibinfo {author} {\bibfnamefont {K.}~\bibnamefont {Rode}},
		\bibinfo {author} {\bibfnamefont {P.}~\bibnamefont {Stamenov}}, \bibinfo
		{author} {\bibfnamefont {J.~M.~D.}\ \bibnamefont {Coey}},\ and\ \bibinfo
		{author} {\bibfnamefont {A.}~\bibnamefont {Brataas}},\ }\href
	{https://doi.org/10.1103/PhysRevB.99.054433} {\bibfield  {journal} {\bibinfo
			{journal} {Phys. Rev. B}\ }\textbf {\bibinfo {volume} {99}},\ \bibinfo
		{pages} {054433} (\bibinfo {year} {2019})}\BibitemShut {NoStop}%
	\bibitem [{\citenamefont {Safin}\ \emph {et~al.}(2020)\citenamefont {Safin},
		\citenamefont {Puliafito}, \citenamefont {Carpentieri}, \citenamefont
		{Finocchio}, \citenamefont {Nikitov}, \citenamefont {Stremoukhov},
		\citenamefont {Kirilyuk}, \citenamefont {Tyberkevych},\ and\ \citenamefont
		{Slavin}}]{Safin2020}%
	\BibitemOpen
	\bibfield  {author} {\bibinfo {author} {\bibfnamefont {A.}~\bibnamefont
			{Safin}}, \bibinfo {author} {\bibfnamefont {V.}~\bibnamefont {Puliafito}},
		\bibinfo {author} {\bibfnamefont {M.}~\bibnamefont {Carpentieri}}, \bibinfo
		{author} {\bibfnamefont {G.}~\bibnamefont {Finocchio}}, \bibinfo {author}
		{\bibfnamefont {S.}~\bibnamefont {Nikitov}}, \bibinfo {author} {\bibfnamefont
			{P.}~\bibnamefont {Stremoukhov}}, \bibinfo {author} {\bibfnamefont
			{A.}~\bibnamefont {Kirilyuk}}, \bibinfo {author} {\bibfnamefont
			{V.}~\bibnamefont {Tyberkevych}},\ and\ \bibinfo {author} {\bibfnamefont
			{A.}~\bibnamefont {Slavin}},\ }\href {https://doi.org/10.1063/5.0031053}
	{\bibfield  {journal} {\bibinfo  {journal} {Appl. Phys. Lett.}\ }\textbf
		{\bibinfo {volume} {117}},\ \bibinfo {pages} {222411} (\bibinfo {year}
		{2020})}\BibitemShut {NoStop}%
	\bibitem [{\citenamefont {Parthasarathy}\ \emph {et~al.}(2021)\citenamefont
		{Parthasarathy}, \citenamefont {Cogulu}, \citenamefont {Kent},\ and\
		\citenamefont {Rakheja}}]{Parthasarathy2021}%
	\BibitemOpen
	\bibfield  {author} {\bibinfo {author} {\bibfnamefont {A.}~\bibnamefont
			{Parthasarathy}}, \bibinfo {author} {\bibfnamefont {E.}~\bibnamefont
			{Cogulu}}, \bibinfo {author} {\bibfnamefont {A.~D.}\ \bibnamefont {Kent}},\
		and\ \bibinfo {author} {\bibfnamefont {S.}~\bibnamefont {Rakheja}},\ }\href
	{https://doi.org/10.1103/PhysRevB.103.024450} {\bibfield  {journal} {\bibinfo
			{journal} {Phys. Rev. B}\ }\textbf {\bibinfo {volume} {103}},\ \bibinfo
		{pages} {024450} (\bibinfo {year} {2021})}\BibitemShut {NoStop}%
	\bibitem [{\citenamefont {Shtanko}\ and\ \citenamefont
		{Prokopenko}(2020)}]{Shtanko2021}%
	\BibitemOpen
	\bibfield  {author} {\bibinfo {author} {\bibfnamefont {O.}~\bibnamefont
			{Shtanko}}\ and\ \bibinfo {author} {\bibfnamefont {O.}~\bibnamefont
			{Prokopenko}},\ }\href {https://doi.org/10.1109/NAP51477.2020.9309712}
	{\bibfield  {journal} {\bibinfo  {journal} {2020 IEEE 10th Int. Conf.
				Nanomater. Appl. Prop.}\ ,\ \bibinfo {pages} {01NMM04}} (\bibinfo {year}
		{2020})}\BibitemShut {NoStop}%
	\bibitem [{\citenamefont {Sulymenko}\ \emph {et~al.}(2018)\citenamefont
		{Sulymenko}, \citenamefont {Prokopenko}, \citenamefont {Lisenkov},
		\citenamefont {{\AA}kerman}, \citenamefont {Tyberkevych}, \citenamefont
		{Slavin},\ and\ \citenamefont {Khymyn}}]{Sulymenko2018}%
	\BibitemOpen
	\bibfield  {author} {\bibinfo {author} {\bibfnamefont {O.}~\bibnamefont
			{Sulymenko}}, \bibinfo {author} {\bibfnamefont {O.}~\bibnamefont
			{Prokopenko}}, \bibinfo {author} {\bibfnamefont {I.}~\bibnamefont
			{Lisenkov}}, \bibinfo {author} {\bibfnamefont {J.}~\bibnamefont
			{{\AA}kerman}}, \bibinfo {author} {\bibfnamefont {V.}~\bibnamefont
			{Tyberkevych}}, \bibinfo {author} {\bibfnamefont {A.~N.}\ \bibnamefont
			{Slavin}},\ and\ \bibinfo {author} {\bibfnamefont {R.}~\bibnamefont
			{Khymyn}},\ }\href {https://doi.org/10.1063/1.5042348} {\bibfield  {journal}
		{\bibinfo  {journal} {J. Appl. Phys.}\ }\textbf {\bibinfo {volume} {124}},\
		\bibinfo {pages} {152115} (\bibinfo {year} {2018})}\BibitemShut {NoStop}%
	\bibitem [{\citenamefont {Khymyn}\ \emph {et~al.}(2018)\citenamefont {Khymyn},
		\citenamefont {Lisenkov}, \citenamefont {Voorheis}, \citenamefont
		{Sulymenko}, \citenamefont {Prokopenko}, \citenamefont {Tiberkevich},
		\citenamefont {Akerman},\ and\ \citenamefont {Slavin}}]{Khymyn2018}%
	\BibitemOpen
	\bibfield  {author} {\bibinfo {author} {\bibfnamefont {R.}~\bibnamefont
			{Khymyn}}, \bibinfo {author} {\bibfnamefont {I.}~\bibnamefont {Lisenkov}},
		\bibinfo {author} {\bibfnamefont {J.}~\bibnamefont {Voorheis}}, \bibinfo
		{author} {\bibfnamefont {O.}~\bibnamefont {Sulymenko}}, \bibinfo {author}
		{\bibfnamefont {O.}~\bibnamefont {Prokopenko}}, \bibinfo {author}
		{\bibfnamefont {V.}~\bibnamefont {Tiberkevich}}, \bibinfo {author}
		{\bibfnamefont {J.}~\bibnamefont {Akerman}},\ and\ \bibinfo {author}
		{\bibfnamefont {A.}~\bibnamefont {Slavin}},\ }\href
	{https://doi.org/10.1038/s41598-018-33697-0} {\bibfield  {journal} {\bibinfo
			{journal} {Sci. Rep.}\ }\textbf {\bibinfo {volume} {8}},\ \bibinfo {pages}
		{15727} (\bibinfo {year} {2018})}\BibitemShut {NoStop}%
	\bibitem [{\citenamefont {Stiles}\ and\ \citenamefont
		{Zangwill}(2002)}]{stiles2002}%
	\BibitemOpen
	\bibfield  {author} {\bibinfo {author} {\bibfnamefont {M.~D.}\ \bibnamefont
			{Stiles}}\ and\ \bibinfo {author} {\bibfnamefont {A.}~\bibnamefont
			{Zangwill}},\ }\href {https://doi.org/10.1103/PhysRevB.66.014407} {\bibfield
		{journal} {\bibinfo  {journal} {Phys. Rev. B}\ }\textbf {\bibinfo {volume}
			{66}},\ \bibinfo {pages} {014407} (\bibinfo {year} {2002})}\BibitemShut
	{NoStop}%
	\bibitem [{\citenamefont {Ralph}\ and\ \citenamefont
		{Stiles}(2008)}]{ralph2008}%
	\BibitemOpen
	\bibfield  {author} {\bibinfo {author} {\bibfnamefont {D.}~\bibnamefont
			{Ralph}}\ and\ \bibinfo {author} {\bibfnamefont {M.}~\bibnamefont {Stiles}},\
	}\href {https://doi.org/10.1016/j.jmmm.2007.12.019} {\bibfield  {journal}
		{\bibinfo  {journal} {J. Magn. Magn. Mater.}\ }\textbf {\bibinfo {volume}
			{320}},\ \bibinfo {pages} {1190} (\bibinfo {year} {2008})}\BibitemShut
	{NoStop}%
	\bibitem [{\citenamefont {Li}\ \emph {et~al.}(2020{\natexlab{a}})\citenamefont
		{Li}, \citenamefont {Wilson}, \citenamefont {Cheng}, \citenamefont {Lohmann},
		\citenamefont {Kavand}, \citenamefont {Yuan}, \citenamefont {Aldosary},
		\citenamefont {Agladze}, \citenamefont {Wei}, \citenamefont {Sherwin},\ and\
		\citenamefont {Shi}}]{Li2020}%
	\BibitemOpen
	\bibfield  {author} {\bibinfo {author} {\bibfnamefont {J.}~\bibnamefont
			{Li}}, \bibinfo {author} {\bibfnamefont {C.~B.}\ \bibnamefont {Wilson}},
		\bibinfo {author} {\bibfnamefont {R.}~\bibnamefont {Cheng}}, \bibinfo
		{author} {\bibfnamefont {M.}~\bibnamefont {Lohmann}}, \bibinfo {author}
		{\bibfnamefont {M.}~\bibnamefont {Kavand}}, \bibinfo {author} {\bibfnamefont
			{W.}~\bibnamefont {Yuan}}, \bibinfo {author} {\bibfnamefont {M.}~\bibnamefont
			{Aldosary}}, \bibinfo {author} {\bibfnamefont {N.}~\bibnamefont {Agladze}},
		\bibinfo {author} {\bibfnamefont {P.}~\bibnamefont {Wei}}, \bibinfo {author}
		{\bibfnamefont {M.~S.}\ \bibnamefont {Sherwin}},\ and\ \bibinfo {author}
		{\bibfnamefont {J.}~\bibnamefont {Shi}},\ }\href
	{https://doi.org/10.1038/s41586-020-1950-4} {\bibfield  {journal} {\bibinfo
			{journal} {Nature}\ }\textbf {\bibinfo {volume} {578}},\ \bibinfo {pages}
		{70} (\bibinfo {year} {2020}{\natexlab{a}})}\BibitemShut {NoStop}%
	\bibitem [{\citenamefont {Vaidya}\ \emph {et~al.}(2020)\citenamefont {Vaidya},
		\citenamefont {Morley}, \citenamefont {{Van Tol}}, \citenamefont {Liu},
		\citenamefont {Cheng}, \citenamefont {Brataas}, \citenamefont {Lederman},\
		and\ \citenamefont {{Del Barco}}}]{Vaidya2020}%
	\BibitemOpen
	\bibfield  {author} {\bibinfo {author} {\bibfnamefont {P.}~\bibnamefont
			{Vaidya}}, \bibinfo {author} {\bibfnamefont {S.~A.}\ \bibnamefont {Morley}},
		\bibinfo {author} {\bibfnamefont {J.}~\bibnamefont {{Van Tol}}}, \bibinfo
		{author} {\bibfnamefont {Y.}~\bibnamefont {Liu}}, \bibinfo {author}
		{\bibfnamefont {R.}~\bibnamefont {Cheng}}, \bibinfo {author} {\bibfnamefont
			{A.}~\bibnamefont {Brataas}}, \bibinfo {author} {\bibfnamefont
			{D.}~\bibnamefont {Lederman}},\ and\ \bibinfo {author} {\bibfnamefont
			{E.}~\bibnamefont {{Del Barco}}},\ }\href
	{https://doi.org/10.1126/science.aaz4247} {\bibfield  {journal} {\bibinfo
			{journal} {Science (80-. ).}\ }\textbf {\bibinfo {volume} {368}},\ \bibinfo
		{pages} {160} (\bibinfo {year} {2020})}\BibitemShut {NoStop}%
	\bibitem [{\citenamefont {Daniels}\ \emph {et~al.}(2018)\citenamefont
		{Daniels}, \citenamefont {Cheng}, \citenamefont {Yu}, \citenamefont {Xiao},\
		and\ \citenamefont {Xiao}}]{Daniels2018}%
	\BibitemOpen
	\bibfield  {author} {\bibinfo {author} {\bibfnamefont {M.~W.}\ \bibnamefont
			{Daniels}}, \bibinfo {author} {\bibfnamefont {R.}~\bibnamefont {Cheng}},
		\bibinfo {author} {\bibfnamefont {W.}~\bibnamefont {Yu}}, \bibinfo {author}
		{\bibfnamefont {J.}~\bibnamefont {Xiao}},\ and\ \bibinfo {author}
		{\bibfnamefont {D.}~\bibnamefont {Xiao}},\ }\href
	{https://doi.org/10.1103/PhysRevB.98.134450} {\bibfield  {journal} {\bibinfo
			{journal} {Phys. Rev. B}\ }\textbf {\bibinfo {volume} {98}},\ \bibinfo
		{pages} {134450} (\bibinfo {year} {2018})}\BibitemShut {NoStop}%
	\bibitem [{\citenamefont {Sulymenko}\ and\ \citenamefont
		{Prokopenko}(2019{\natexlab{a}})}]{Sulymenko2019}%
	\BibitemOpen
	\bibfield  {author} {\bibinfo {author} {\bibfnamefont {O.~R.}\ \bibnamefont
			{Sulymenko}}\ and\ \bibinfo {author} {\bibfnamefont {O.~V.}\ \bibnamefont
			{Prokopenko}},\ }\href {https://doi.org/10.1109/CAOL46282.2019.9019409}
	{\bibfield  {journal} {\bibinfo  {journal} {Proc. Int. Conf. Adv.
				Optoelectron. Lasers, CAOL}\ }\textbf {\bibinfo {volume} {2019-Septe}},\
		\bibinfo {pages} {533} (\bibinfo {year} {2019}{\natexlab{a}})}\BibitemShut
	{NoStop}%
	\bibitem [{\citenamefont {Sulymenko}\ and\ \citenamefont
		{Prokopenko}(2019{\natexlab{b}})}]{Sulymenko2019a}%
	\BibitemOpen
	\bibfield  {author} {\bibinfo {author} {\bibfnamefont {O.~R.}\ \bibnamefont
			{Sulymenko}}\ and\ \bibinfo {author} {\bibfnamefont {O.~V.}\ \bibnamefont
			{Prokopenko}},\ }\href {https://doi.org/10.1109/ELNANO.2019.8783266}
	{\bibfield  {journal} {\bibinfo  {journal} {2019 IEEE 39th Int. Conf.
				Electron. Nanotechnology, ELNANO 2019 - Proc.}\ ,\ \bibinfo {pages} {132}}
		(\bibinfo {year} {2019}{\natexlab{b}})}\BibitemShut {NoStop}%
	\bibitem [{\citenamefont {Grollier}\ \emph {et~al.}(2020)\citenamefont
		{Grollier}, \citenamefont {Querlioz}, \citenamefont {Camsari}, \citenamefont
		{Everschor-Sitte}, \citenamefont {Fukami},\ and\ \citenamefont
		{Stiles}}]{Grollier2020}%
	\BibitemOpen
	\bibfield  {author} {\bibinfo {author} {\bibfnamefont {J.}~\bibnamefont
			{Grollier}}, \bibinfo {author} {\bibfnamefont {D.}~\bibnamefont {Querlioz}},
		\bibinfo {author} {\bibfnamefont {K.~Y.}\ \bibnamefont {Camsari}}, \bibinfo
		{author} {\bibfnamefont {K.}~\bibnamefont {Everschor-Sitte}}, \bibinfo
		{author} {\bibfnamefont {S.}~\bibnamefont {Fukami}},\ and\ \bibinfo {author}
		{\bibfnamefont {M.~D.}\ \bibnamefont {Stiles}},\ }\href
	{https://doi.org/10.1038/s41928-019-0360-9} {\bibinfo {title} {{Neuromorphic
				spintronics}}} (\bibinfo {year} {2020})\BibitemShut {NoStop}%
	\bibitem [{\citenamefont {Chen}\ \emph {et~al.}(2016)\citenamefont {Chen},
		\citenamefont {Dumas}, \citenamefont {Eklund}, \citenamefont {Muduli},
		\citenamefont {Houshang}, \citenamefont {Awad}, \citenamefont
		{D{\"{u}}rrenfeld}, \citenamefont {Malm}, \citenamefont {Rusu},\ and\
		\citenamefont {Akerman}}]{Chen2016}%
	\BibitemOpen
	\bibfield  {author} {\bibinfo {author} {\bibfnamefont {T.}~\bibnamefont
			{Chen}}, \bibinfo {author} {\bibfnamefont {R.~K.}\ \bibnamefont {Dumas}},
		\bibinfo {author} {\bibfnamefont {A.}~\bibnamefont {Eklund}}, \bibinfo
		{author} {\bibfnamefont {P.~K.}\ \bibnamefont {Muduli}}, \bibinfo {author}
		{\bibfnamefont {A.}~\bibnamefont {Houshang}}, \bibinfo {author}
		{\bibfnamefont {A.~A.}\ \bibnamefont {Awad}}, \bibinfo {author}
		{\bibfnamefont {P.}~\bibnamefont {D{\"{u}}rrenfeld}}, \bibinfo {author}
		{\bibfnamefont {B.~G.}\ \bibnamefont {Malm}}, \bibinfo {author}
		{\bibfnamefont {A.}~\bibnamefont {Rusu}},\ and\ \bibinfo {author}
		{\bibfnamefont {J.}~\bibnamefont {Akerman}},\ }\href
	{https://doi.org/10.1109/JPROC.2016.2554518} {\bibfield  {journal} {\bibinfo
			{journal} {Proc. IEEE}\ }\textbf {\bibinfo {volume} {104}},\ \bibinfo {pages}
		{1919} (\bibinfo {year} {2016})}\BibitemShut {NoStop}%
	\bibitem [{\citenamefont {Demidov}\ \emph {et~al.}(2017)\citenamefont
		{Demidov}, \citenamefont {Urazhdin}, \citenamefont {de~Loubens},
		\citenamefont {Klein}, \citenamefont {Cros}, \citenamefont {Anane},\ and\
		\citenamefont {Demokritov}}]{Demidov2017}%
	\BibitemOpen
	\bibfield  {author} {\bibinfo {author} {\bibfnamefont {V.}~\bibnamefont
			{Demidov}}, \bibinfo {author} {\bibfnamefont {S.}~\bibnamefont {Urazhdin}},
		\bibinfo {author} {\bibfnamefont {G.}~\bibnamefont {de~Loubens}}, \bibinfo
		{author} {\bibfnamefont {O.}~\bibnamefont {Klein}}, \bibinfo {author}
		{\bibfnamefont {V.}~\bibnamefont {Cros}}, \bibinfo {author} {\bibfnamefont
			{A.}~\bibnamefont {Anane}},\ and\ \bibinfo {author} {\bibfnamefont
			{S.}~\bibnamefont {Demokritov}},\ }\href
	{https://doi.org/10.1016/J.PHYSREP.2017.01.001} {\bibfield  {journal}
		{\bibinfo  {journal} {Phys. Rep.}\ }\textbf {\bibinfo {volume} {673}},\
		\bibinfo {pages} {1} (\bibinfo {year} {2017})}\BibitemShut {NoStop}%
	\bibitem [{\citenamefont {Hoefer}\ \emph {et~al.}(2010)\citenamefont {Hoefer},
		\citenamefont {Silva},\ and\ \citenamefont {Keller}}]{Hoefer2010}%
	\BibitemOpen
	\bibfield  {author} {\bibinfo {author} {\bibfnamefont {M.~A.}\ \bibnamefont
			{Hoefer}}, \bibinfo {author} {\bibfnamefont {T.~J.}\ \bibnamefont {Silva}},\
		and\ \bibinfo {author} {\bibfnamefont {M.~W.}\ \bibnamefont {Keller}},\
	}\href {https://doi.org/10.1103/PHYSREVB.82.054432/FIGURES/12/MEDIUM}
	{\bibfield  {journal} {\bibinfo  {journal} {Phys. Rev. B}\ }\textbf {\bibinfo
			{volume} {82}},\ \bibinfo {pages} {054432} (\bibinfo {year}
		{2010})}\BibitemShut {NoStop}%
	\bibitem [{\citenamefont {Mohseni}\ \emph {et~al.}(2013)\citenamefont
		{Mohseni}, \citenamefont {Sani}, \citenamefont {Persson}, \citenamefont {{Anh
				Nguyen}}, \citenamefont {Chung}, \citenamefont {Pogoryelov}, \citenamefont
		{Muduli}, \citenamefont {Iacocca}, \citenamefont {Eklund}, \citenamefont
		{Dumas}, \citenamefont {Bonetti}, \citenamefont {Deac}, \citenamefont
		{Hoefer},\ and\ \citenamefont {{\AA}kerman}}]{Mohseni2013}%
	\BibitemOpen
	\bibfield  {author} {\bibinfo {author} {\bibfnamefont {S.~M.}\ \bibnamefont
			{Mohseni}}, \bibinfo {author} {\bibfnamefont {S.~R.}\ \bibnamefont {Sani}},
		\bibinfo {author} {\bibfnamefont {J.}~\bibnamefont {Persson}}, \bibinfo
		{author} {\bibfnamefont {T.~N.}\ \bibnamefont {{Anh Nguyen}}}, \bibinfo
		{author} {\bibfnamefont {S.}~\bibnamefont {Chung}}, \bibinfo {author}
		{\bibfnamefont {Y.}~\bibnamefont {Pogoryelov}}, \bibinfo {author}
		{\bibfnamefont {P.~K.}\ \bibnamefont {Muduli}}, \bibinfo {author}
		{\bibfnamefont {E.}~\bibnamefont {Iacocca}}, \bibinfo {author} {\bibfnamefont
			{A.}~\bibnamefont {Eklund}}, \bibinfo {author} {\bibfnamefont {R.~K.}\
			\bibnamefont {Dumas}}, \bibinfo {author} {\bibfnamefont {S.}~\bibnamefont
			{Bonetti}}, \bibinfo {author} {\bibfnamefont {A.}~\bibnamefont {Deac}},
		\bibinfo {author} {\bibfnamefont {M.~A.}\ \bibnamefont {Hoefer}},\ and\
		\bibinfo {author} {\bibfnamefont {J.}~\bibnamefont {{\AA}kerman}},\ }\href
	{https://doi.org/10.1126/SCIENCE.1230155/SUPPL_FILE/MOHESNI-SOM.PDF}
	{\bibfield  {journal} {\bibinfo  {journal} {Science (80-. ).}\ }\textbf
		{\bibinfo {volume} {339}},\ \bibinfo {pages} {1295} (\bibinfo {year}
		{2013})}\BibitemShut {NoStop}%
	\bibitem [{\citenamefont {Mohseni}\ \emph
		{et~al.}(2018{\natexlab{a}})\citenamefont {Mohseni}, \citenamefont {Hamdi},
		\citenamefont {Yazdi}, \citenamefont {Banuazizi}, \citenamefont {Chung},
		\citenamefont {Sani}, \citenamefont {{\AA}kerman},\ and\ \citenamefont
		{Mohseni}}]{Mohseni2018}%
	\BibitemOpen
	\bibfield  {author} {\bibinfo {author} {\bibfnamefont {M.}~\bibnamefont
			{Mohseni}}, \bibinfo {author} {\bibfnamefont {M.}~\bibnamefont {Hamdi}},
		\bibinfo {author} {\bibfnamefont {H.~F.}\ \bibnamefont {Yazdi}}, \bibinfo
		{author} {\bibfnamefont {S.~A.~H.}\ \bibnamefont {Banuazizi}}, \bibinfo
		{author} {\bibfnamefont {S.}~\bibnamefont {Chung}}, \bibinfo {author}
		{\bibfnamefont {S.~R.}\ \bibnamefont {Sani}}, \bibinfo {author}
		{\bibfnamefont {J.}~\bibnamefont {{\AA}kerman}},\ and\ \bibinfo {author}
		{\bibfnamefont {M.}~\bibnamefont {Mohseni}},\ }\href
	{https://doi.org/10.1103/PhysRevB.97.184402} {\bibfield  {journal} {\bibinfo
			{journal} {Phys. Rev. B}\ }\textbf {\bibinfo {volume} {97}},\ \bibinfo
		{pages} {184402} (\bibinfo {year} {2018}{\natexlab{a}})}\BibitemShut
	{NoStop}%
	\bibitem [{\citenamefont {Slonczewski}(1996)}]{slonczewski1996}%
	\BibitemOpen
	\bibfield  {author} {\bibinfo {author} {\bibfnamefont {J.}~\bibnamefont
			{Slonczewski}},\ }\href {https://doi.org/10.1016/0304-8853(96)00062-5}
	{\bibfield  {journal} {\bibinfo  {journal} {J. Magn. Magn. Mater.}\ }\textbf
		{\bibinfo {volume} {159}},\ \bibinfo {pages} {L1} (\bibinfo {year}
		{1996})}\BibitemShut {NoStop}%
	\bibitem [{\citenamefont {Slonczewski}(1999)}]{Slonczewski1999}%
	\BibitemOpen
	\bibfield  {author} {\bibinfo {author} {\bibfnamefont {J.~C.}\ \bibnamefont
			{Slonczewski}},\ }\href {https://doi.org/10.1016/S0304-8853(99)00043-8}
	{\bibfield  {journal} {\bibinfo  {journal} {J. Magn. Magn. Mater.}\ }\textbf
		{\bibinfo {volume} {195}},\ \bibinfo {pages} {L261} (\bibinfo {year}
		{1999})}\BibitemShut {NoStop}%
	\bibitem [{\citenamefont {Hoefer}\ \emph {et~al.}(2005)\citenamefont {Hoefer},
		\citenamefont {Ablowitz}, \citenamefont {Ilan}, \citenamefont {Pufall},\ and\
		\citenamefont {Silva}}]{Hoefer2005}%
	\BibitemOpen
	\bibfield  {author} {\bibinfo {author} {\bibfnamefont {M.~A.}\ \bibnamefont
			{Hoefer}}, \bibinfo {author} {\bibfnamefont {M.~J.}\ \bibnamefont
			{Ablowitz}}, \bibinfo {author} {\bibfnamefont {B.}~\bibnamefont {Ilan}},
		\bibinfo {author} {\bibfnamefont {M.~R.}\ \bibnamefont {Pufall}},\ and\
		\bibinfo {author} {\bibfnamefont {T.~J.}\ \bibnamefont {Silva}},\ }\href
	{https://doi.org/10.1103/PhysRevLett.95.267206} {\bibfield  {journal}
		{\bibinfo  {journal} {Phys. Rev. Lett.}\ }\textbf {\bibinfo {volume} {95}},\
		\bibinfo {pages} {267206} (\bibinfo {year} {2005})}\BibitemShut {NoStop}%
	\bibitem [{\citenamefont {Mohseni}\ \emph
		{et~al.}(2018{\natexlab{b}})\citenamefont {Mohseni}, \citenamefont {Yazdi},
		\citenamefont {Hamdi}, \citenamefont {Br{\"{a}}cher},\ and\ \citenamefont
		{Mohseni}}]{Mohseni2018b}%
	\BibitemOpen
	\bibfield  {author} {\bibinfo {author} {\bibfnamefont {S.~M.}\ \bibnamefont
			{Mohseni}}, \bibinfo {author} {\bibfnamefont {H.~F.}\ \bibnamefont {Yazdi}},
		\bibinfo {author} {\bibfnamefont {M.}~\bibnamefont {Hamdi}}, \bibinfo
		{author} {\bibfnamefont {T.}~\bibnamefont {Br{\"{a}}cher}},\ and\ \bibinfo
		{author} {\bibfnamefont {S.~M.}\ \bibnamefont {Mohseni}},\ }\href
	{https://doi.org/10.1016/J.JMMM.2017.09.032} {\bibfield  {journal} {\bibinfo
			{journal} {J. Magn. Magn. Mater.}\ }\textbf {\bibinfo {volume} {450}},\
		\bibinfo {pages} {40} (\bibinfo {year} {2018}{\natexlab{b}})}\BibitemShut
	{NoStop}%
	\bibitem [{\citenamefont {Kendziorczyk}\ \emph {et~al.}(2014)\citenamefont
		{Kendziorczyk}, \citenamefont {Demokritov},\ and\ \citenamefont
		{Kuhn}}]{Kendziorczyk2014}%
	\BibitemOpen
	\bibfield  {author} {\bibinfo {author} {\bibfnamefont {T.}~\bibnamefont
			{Kendziorczyk}}, \bibinfo {author} {\bibfnamefont {S.~O.}\ \bibnamefont
			{Demokritov}},\ and\ \bibinfo {author} {\bibfnamefont {T.}~\bibnamefont
			{Kuhn}},\ }\href {https://doi.org/10.1103/PhysRevB.90.054414} {\bibfield
		{journal} {\bibinfo  {journal} {Phys. Rev. B}\ }\textbf {\bibinfo {volume}
			{90}},\ \bibinfo {pages} {054414} (\bibinfo {year} {2014})}\BibitemShut
	{NoStop}%
	\bibitem [{\citenamefont {Houshang}\ \emph {et~al.}(2016)\citenamefont
		{Houshang}, \citenamefont {Iacocca}, \citenamefont {D{\"{u}}rrenfeld},
		\citenamefont {Sani}, \citenamefont {{\AA}kerman},\ and\ \citenamefont
		{Dumas}}]{Houshang2016}%
	\BibitemOpen
	\bibfield  {author} {\bibinfo {author} {\bibfnamefont {A.}~\bibnamefont
			{Houshang}}, \bibinfo {author} {\bibfnamefont {E.}~\bibnamefont {Iacocca}},
		\bibinfo {author} {\bibfnamefont {P.}~\bibnamefont {D{\"{u}}rrenfeld}},
		\bibinfo {author} {\bibfnamefont {S.~R.}\ \bibnamefont {Sani}}, \bibinfo
		{author} {\bibfnamefont {J.}~\bibnamefont {{\AA}kerman}},\ and\ \bibinfo
		{author} {\bibfnamefont {R.~K.}\ \bibnamefont {Dumas}},\ }\href
	{https://doi.org/10.1038/nnano.2015.280} {\bibfield  {journal} {\bibinfo
			{journal} {Nat. Nanotechnol.}\ }\textbf {\bibinfo {volume} {11}},\ \bibinfo
		{pages} {280} (\bibinfo {year} {2016})}\BibitemShut {NoStop}%
	\bibitem [{\citenamefont {Kendziorczyk}\ and\ \citenamefont
		{Kuhn}(2016)}]{Kendziorczyk2016}%
	\BibitemOpen
	\bibfield  {author} {\bibinfo {author} {\bibfnamefont {T.}~\bibnamefont
			{Kendziorczyk}}\ and\ \bibinfo {author} {\bibfnamefont {T.}~\bibnamefont
			{Kuhn}},\ }\href {https://doi.org/10.1103/PhysRevB.93.134413} {\bibfield
		{journal} {\bibinfo  {journal} {Phys. Rev. B}\ }\textbf {\bibinfo {volume}
			{93}},\ \bibinfo {pages} {134413} (\bibinfo {year} {2016})}\BibitemShut
	{NoStop}%
	\bibitem [{\citenamefont {Awad}\ \emph {et~al.}(2017)\citenamefont {Awad},
		\citenamefont {D{\"{u}}rrenfeld}, \citenamefont {Houshang}, \citenamefont
		{Dvornik}, \citenamefont {Iacocca}, \citenamefont {Dumas},\ and\
		\citenamefont {{\AA}kerman}}]{Awad2017}%
	\BibitemOpen
	\bibfield  {author} {\bibinfo {author} {\bibfnamefont {A.~A.}\ \bibnamefont
			{Awad}}, \bibinfo {author} {\bibfnamefont {P.}~\bibnamefont
			{D{\"{u}}rrenfeld}}, \bibinfo {author} {\bibfnamefont {A.}~\bibnamefont
			{Houshang}}, \bibinfo {author} {\bibfnamefont {M.}~\bibnamefont {Dvornik}},
		\bibinfo {author} {\bibfnamefont {E.}~\bibnamefont {Iacocca}}, \bibinfo
		{author} {\bibfnamefont {R.~K.}\ \bibnamefont {Dumas}},\ and\ \bibinfo
		{author} {\bibfnamefont {J.}~\bibnamefont {{\AA}kerman}},\ }\href
	{https://doi.org/10.1038/nphys3927} {\bibfield  {journal} {\bibinfo
			{journal} {Nat. Phys.}\ }\textbf {\bibinfo {volume} {13}},\ \bibinfo {pages}
		{292} (\bibinfo {year} {2017})}\BibitemShut {NoStop}%
	\bibitem [{\citenamefont {Zahedinejad}\ \emph {et~al.}(2020)\citenamefont
		{Zahedinejad}, \citenamefont {Awad}, \citenamefont {Muralidhar},
		\citenamefont {Khymyn}, \citenamefont {Fulara}, \citenamefont {Mazraati},
		\citenamefont {Dvornik},\ and\ \citenamefont
		{{\AA}kerman}}]{Zahedinejad2020}%
	\BibitemOpen
	\bibfield  {author} {\bibinfo {author} {\bibfnamefont {M.}~\bibnamefont
			{Zahedinejad}}, \bibinfo {author} {\bibfnamefont {A.~A.}\ \bibnamefont
			{Awad}}, \bibinfo {author} {\bibfnamefont {S.}~\bibnamefont {Muralidhar}},
		\bibinfo {author} {\bibfnamefont {R.}~\bibnamefont {Khymyn}}, \bibinfo
		{author} {\bibfnamefont {H.}~\bibnamefont {Fulara}}, \bibinfo {author}
		{\bibfnamefont {H.}~\bibnamefont {Mazraati}}, \bibinfo {author}
		{\bibfnamefont {M.}~\bibnamefont {Dvornik}},\ and\ \bibinfo {author}
		{\bibfnamefont {J.}~\bibnamefont {{\AA}kerman}},\ }\href
	{https://doi.org/10.1038/s41565-019-0593-9} {\bibfield  {journal} {\bibinfo
			{journal} {Nat. Nanotechnol.}\ }\textbf {\bibinfo {volume} {15}},\ \bibinfo
		{pages} {47} (\bibinfo {year} {2020})}\BibitemShut {NoStop}%
	\bibitem [{\citenamefont {Zahedinejad}\ \emph {et~al.}(2022)\citenamefont
		{Zahedinejad}, \citenamefont {Fulara}, \citenamefont {Khymyn}, \citenamefont
		{Houshang}, \citenamefont {Dvornik}, \citenamefont {Fukami}, \citenamefont
		{Kanai}, \citenamefont {Ohno},\ and\ \citenamefont
		{{\AA}kerman}}]{Zahedinejad2022}%
	\BibitemOpen
	\bibfield  {author} {\bibinfo {author} {\bibfnamefont {M.}~\bibnamefont
			{Zahedinejad}}, \bibinfo {author} {\bibfnamefont {H.}~\bibnamefont {Fulara}},
		\bibinfo {author} {\bibfnamefont {R.}~\bibnamefont {Khymyn}}, \bibinfo
		{author} {\bibfnamefont {A.}~\bibnamefont {Houshang}}, \bibinfo {author}
		{\bibfnamefont {M.}~\bibnamefont {Dvornik}}, \bibinfo {author} {\bibfnamefont
			{S.}~\bibnamefont {Fukami}}, \bibinfo {author} {\bibfnamefont
			{S.}~\bibnamefont {Kanai}}, \bibinfo {author} {\bibfnamefont
			{H.}~\bibnamefont {Ohno}},\ and\ \bibinfo {author} {\bibfnamefont
			{J.}~\bibnamefont {{\AA}kerman}},\ }\href
	{https://doi.org/10.1038/s41563-021-01153-6} {\bibfield  {journal} {\bibinfo
			{journal} {Nat. Mater.}\ }\textbf {\bibinfo {volume} {21}},\ \bibinfo {pages}
		{81} (\bibinfo {year} {2022})}\BibitemShut {NoStop}%
	\bibitem [{\citenamefont {Berger}(1996)}]{berger1996}%
	\BibitemOpen
	\bibfield  {author} {\bibinfo {author} {\bibfnamefont {L.}~\bibnamefont
			{Berger}},\ }\href {https://doi.org/10.1103/PhysRevB.54.9353} {\bibfield
		{journal} {\bibinfo  {journal} {Phys. Rev. B}\ }\textbf {\bibinfo {volume}
			{54}},\ \bibinfo {pages} {9353} (\bibinfo {year} {1996})}\BibitemShut
	{NoStop}%
	\bibitem [{\citenamefont {Chung}\ \emph {et~al.}(2016)\citenamefont {Chung},
		\citenamefont {Eklund}, \citenamefont {Iacocca}, \citenamefont {Mohseni},
		\citenamefont {Sani}, \citenamefont {Bookman}, \citenamefont {Hoefer},
		\citenamefont {Dumas},\ and\ \citenamefont {{\AA}kerman}}]{Chung2016}%
	\BibitemOpen
	\bibfield  {author} {\bibinfo {author} {\bibfnamefont {S.}~\bibnamefont
			{Chung}}, \bibinfo {author} {\bibfnamefont {A.}~\bibnamefont {Eklund}},
		\bibinfo {author} {\bibfnamefont {E.}~\bibnamefont {Iacocca}}, \bibinfo
		{author} {\bibfnamefont {S.~M.}\ \bibnamefont {Mohseni}}, \bibinfo {author}
		{\bibfnamefont {S.~R.}\ \bibnamefont {Sani}}, \bibinfo {author}
		{\bibfnamefont {L.}~\bibnamefont {Bookman}}, \bibinfo {author} {\bibfnamefont
			{M.~A.}\ \bibnamefont {Hoefer}}, \bibinfo {author} {\bibfnamefont {R.~K.}\
			\bibnamefont {Dumas}},\ and\ \bibinfo {author} {\bibfnamefont
			{J.}~\bibnamefont {{\AA}kerman}},\ }\href
	{https://doi.org/10.1038/ncomms11209} {\bibfield  {journal} {\bibinfo
			{journal} {Nat. Commun.}\ }\textbf {\bibinfo {volume} {7}},\ \bibinfo {pages}
		{1} (\bibinfo {year} {2016})}\BibitemShut {NoStop}%
	\bibitem [{\citenamefont {Hals}\ \emph {et~al.}(2011)\citenamefont {Hals},
		\citenamefont {Tserkovnyak},\ and\ \citenamefont {Brataas}}]{Hals2011}%
	\BibitemOpen
	\bibfield  {author} {\bibinfo {author} {\bibfnamefont {K.~M.~D.}\
			\bibnamefont {Hals}}, \bibinfo {author} {\bibfnamefont {Y.}~\bibnamefont
			{Tserkovnyak}},\ and\ \bibinfo {author} {\bibfnamefont {A.}~\bibnamefont
			{Brataas}},\ }\href {https://doi.org/10.1103/PhysRevLett.106.107206}
	{\bibfield  {journal} {\bibinfo  {journal} {Phys. Rev. Lett.}\ }\textbf
		{\bibinfo {volume} {106}},\ \bibinfo {pages} {107206} (\bibinfo {year}
		{2011})}\BibitemShut {NoStop}%
	\bibitem [{\citenamefont {Galkina}\ and\ \citenamefont
		{Ivanov}(2018)}]{Galkina2018}%
	\BibitemOpen
	\bibfield  {author} {\bibinfo {author} {\bibfnamefont {E.~G.}\ \bibnamefont
			{Galkina}}\ and\ \bibinfo {author} {\bibfnamefont {B.~A.}\ \bibnamefont
			{Ivanov}},\ }\href {https://doi.org/10.1063/1.5041427} {\bibfield  {journal}
		{\bibinfo  {journal} {Low Temp. Phys.}\ }\textbf {\bibinfo {volume} {44}},\
		\bibinfo {pages} {618} (\bibinfo {year} {2018})}\BibitemShut {NoStop}%
	\bibitem [{\citenamefont {Wang}\ \emph {et~al.}(2021)\citenamefont {Wang},
		\citenamefont {Nie}, \citenamefont {Chotorlishvili}, \citenamefont {Xia},
		\citenamefont {Berakdar},\ and\ \citenamefont {Guo}}]{Wang2021}%
	\BibitemOpen
	\bibfield  {author} {\bibinfo {author} {\bibfnamefont {X.-g.}\ \bibnamefont
			{Wang}}, \bibinfo {author} {\bibfnamefont {Y.-Z.}\ \bibnamefont {Nie}},
		\bibinfo {author} {\bibfnamefont {L.}~\bibnamefont {Chotorlishvili}},
		\bibinfo {author} {\bibfnamefont {Q.-l.}\ \bibnamefont {Xia}}, \bibinfo
		{author} {\bibfnamefont {J.}~\bibnamefont {Berakdar}},\ and\ \bibinfo
		{author} {\bibfnamefont {G.-h.}\ \bibnamefont {Guo}},\ }\href
	{https://doi.org/10.1103/PhysRevB.103.064404} {\bibfield  {journal} {\bibinfo
			{journal} {Phys. Rev. B}\ }\textbf {\bibinfo {volume} {103}},\ \bibinfo
		{pages} {064404} (\bibinfo {year} {2021})}\BibitemShut {NoStop}%
	\bibitem [{Sup()}]{SupplementalNote1}%
	\BibitemOpen
	\href@noop {} {}\bibinfo {note} {See Supplemental Material at (link) for derivations of the SW equation as well as the solution to SW equation and derivation of the current density and frequency at the threshold in section S-I. The derivations of continuum model free energy from micromagnetic energy terms used in the employed micromagnetic simulation package (OOMMF) \cite{Donahue1999} is presented in section S-II. We employ a similar approach to Ref. \onlinecite{Li2020a} in section S-II. However, by correctly considering the exchange term for a pair of spins and appropriate choice of the coordinate system we do resolve the issue of manually plugging a factor of 1/2 in Ref. \onlinecite{Li2020a}. See section S-III for the details of the micromagnetic simulations. We also present a comparative study of SW dispersion of an AFM obtained by OOMMF and a two sublattice micromagnetic model (Boris Computational Spintronics \cite{Lepadatu2020}) as a validation tool for our micromagnetic simulations. See supplemental video 1 and 2 for the visualization of the Slonczewski spin wave obtained from micromagnetic simulations at the threshold and above threshold, respectively. Derivation of Eq. \ref{SpinPumping} is presented in section S-IV. The scientific colour maps developed by Crameri \textit{et. al.} \cite{Crameri2021} is used in this study to prevent visual distortion of the data and exclusion of readers with colour-vision deficiencies \cite{Crameri2020}.}\BibitemShut {Stop}%
	\bibitem [{\citenamefont {Donahue}\ and\ \citenamefont
		{Porter}(1999)}]{Donahue1999}%
	\BibitemOpen
	\bibfield  {author} {\bibinfo {author} {\bibfnamefont {M.~J.}\ \bibnamefont
			{Donahue}}\ and\ \bibinfo {author} {\bibfnamefont {D.~G.}\ \bibnamefont
			{Porter}}\ }\href {https://doi.org/10.6028/NIST.IR.6376}
	{10.6028/NIST.IR.6376} (\bibinfo {year} {1999})\BibitemShut {NoStop}%
	\bibitem [{\citenamefont {Venkat}\ \emph {et~al.}(2018)\citenamefont {Venkat},
		\citenamefont {Fangohr},\ and\ \citenamefont {Prabhakar}}]{Venkat2018}%
	\BibitemOpen
	\bibfield  {author} {\bibinfo {author} {\bibfnamefont {G.}~\bibnamefont
			{Venkat}}, \bibinfo {author} {\bibfnamefont {H.}~\bibnamefont {Fangohr}},\
		and\ \bibinfo {author} {\bibfnamefont {A.}~\bibnamefont {Prabhakar}},\ }\href
	{https://doi.org/10.1016/J.JMMM.2017.06.057} {\bibfield  {journal} {\bibinfo
			{journal} {J. Magn. Magn. Mater.}\ }\textbf {\bibinfo {volume} {450}},\
		\bibinfo {pages} {34} (\bibinfo {year} {2018})}\BibitemShut {NoStop}%
	\bibitem [{\citenamefont {Foner}(1963)}]{Foner1963}%
	\BibitemOpen
	\bibfield  {author} {\bibinfo {author} {\bibfnamefont {S.}~\bibnamefont
			{Foner}},\ }\href {https://doi.org/10.1103/PhysRev.130.183} {\bibfield
		{journal} {\bibinfo  {journal} {Phys. Rev.}\ }\textbf {\bibinfo {volume}
			{130}},\ \bibinfo {pages} {183} (\bibinfo {year} {1963})}\BibitemShut
	{NoStop}%
	\bibitem [{\citenamefont {Artman}\ \emph {et~al.}(1965)\citenamefont {Artman},
		\citenamefont {Murphy},\ and\ \citenamefont {Foner}}]{Artman1965}%
	\BibitemOpen
	\bibfield  {author} {\bibinfo {author} {\bibfnamefont {J.~O.}\ \bibnamefont
			{Artman}}, \bibinfo {author} {\bibfnamefont {J.~C.}\ \bibnamefont {Murphy}},\
		and\ \bibinfo {author} {\bibfnamefont {S.}~\bibnamefont {Foner}},\ }\href
	{https://doi.org/10.1103/PhysRev.138.A912} {\bibfield  {journal} {\bibinfo
			{journal} {Phys. Rev.}\ }\textbf {\bibinfo {volume} {138}},\ \bibinfo {pages}
		{A912} (\bibinfo {year} {1965})}\BibitemShut {NoStop}%
	\bibitem [{\citenamefont {Saidl}\ \emph {et~al.}(2017)\citenamefont {Saidl},
		\citenamefont {N{\v{e}}mec}, \citenamefont {Wadley}, \citenamefont {Hills},
		\citenamefont {Campion}, \citenamefont {Nov{\'{a}}k}, \citenamefont
		{Edmonds}, \citenamefont {Maccherozzi}, \citenamefont {Dhesi}, \citenamefont
		{Gallagher}, \citenamefont {Troj{\'{a}}nek}, \citenamefont {Kune{\v{s}}},
		\citenamefont {{\v{Z}}elezn{\'{y}}}, \citenamefont {Mal{\'{y}}},\ and\
		\citenamefont {Jungwirth}}]{Saidl2017}%
	\BibitemOpen
	\bibfield  {author} {\bibinfo {author} {\bibfnamefont {V.}~\bibnamefont
			{Saidl}}, \bibinfo {author} {\bibfnamefont {P.}~\bibnamefont {N{\v{e}}mec}},
		\bibinfo {author} {\bibfnamefont {P.}~\bibnamefont {Wadley}}, \bibinfo
		{author} {\bibfnamefont {V.}~\bibnamefont {Hills}}, \bibinfo {author}
		{\bibfnamefont {R.~P.}\ \bibnamefont {Campion}}, \bibinfo {author}
		{\bibfnamefont {V.}~\bibnamefont {Nov{\'{a}}k}}, \bibinfo {author}
		{\bibfnamefont {K.~W.}\ \bibnamefont {Edmonds}}, \bibinfo {author}
		{\bibfnamefont {F.}~\bibnamefont {Maccherozzi}}, \bibinfo {author}
		{\bibfnamefont {S.~S.}\ \bibnamefont {Dhesi}}, \bibinfo {author}
		{\bibfnamefont {B.~L.}\ \bibnamefont {Gallagher}}, \bibinfo {author}
		{\bibfnamefont {F.}~\bibnamefont {Troj{\'{a}}nek}}, \bibinfo {author}
		{\bibfnamefont {J.}~\bibnamefont {Kune{\v{s}}}}, \bibinfo {author}
		{\bibfnamefont {J.}~\bibnamefont {{\v{Z}}elezn{\'{y}}}}, \bibinfo {author}
		{\bibfnamefont {P.}~\bibnamefont {Mal{\'{y}}}},\ and\ \bibinfo {author}
		{\bibfnamefont {T.}~\bibnamefont {Jungwirth}},\ }\href
	{https://doi.org/10.1038/nphoton.2016.255} {\bibfield  {journal} {\bibinfo
			{journal} {Nat. Photonics}\ }\textbf {\bibinfo {volume} {11}},\ \bibinfo
		{pages} {91} (\bibinfo {year} {2017})}\BibitemShut {NoStop}%
	\bibitem [{\citenamefont {Lebrun}\ \emph {et~al.}(2018)\citenamefont {Lebrun},
		\citenamefont {Ross}, \citenamefont {Bender}, \citenamefont {Qaiumzadeh},
		\citenamefont {Baldrati}, \citenamefont {Cramer}, \citenamefont {Brataas},
		\citenamefont {Duine},\ and\ \citenamefont {Kl{\"{a}}ui}}]{Lebrun2018}%
	\BibitemOpen
	\bibfield  {author} {\bibinfo {author} {\bibfnamefont {R.}~\bibnamefont
			{Lebrun}}, \bibinfo {author} {\bibfnamefont {A.}~\bibnamefont {Ross}},
		\bibinfo {author} {\bibfnamefont {S.~A.}\ \bibnamefont {Bender}}, \bibinfo
		{author} {\bibfnamefont {A.}~\bibnamefont {Qaiumzadeh}}, \bibinfo {author}
		{\bibfnamefont {L.}~\bibnamefont {Baldrati}}, \bibinfo {author}
		{\bibfnamefont {J.}~\bibnamefont {Cramer}}, \bibinfo {author} {\bibfnamefont
			{A.}~\bibnamefont {Brataas}}, \bibinfo {author} {\bibfnamefont {R.~A.}\
			\bibnamefont {Duine}},\ and\ \bibinfo {author} {\bibfnamefont
			{M.}~\bibnamefont {Kl{\"{a}}ui}},\ }\href
	{https://doi.org/10.1038/s41586-018-0490-7} {\bibfield  {journal} {\bibinfo
			{journal} {Nature}\ }\textbf {\bibinfo {volume} {561}},\ \bibinfo {pages}
		{222} (\bibinfo {year} {2018})}\BibitemShut {NoStop}%
	\bibitem [{\citenamefont {Wang}\ \emph {et~al.}(2020)\citenamefont {Wang},
		\citenamefont {Andrews}, \citenamefont {Reimers}, \citenamefont {Amin},
		\citenamefont {Wadley}, \citenamefont {Campion}, \citenamefont {Poole},
		\citenamefont {Felton}, \citenamefont {Edmonds}, \citenamefont {Gallagher},
		\citenamefont {Rushforth}, \citenamefont {Makarovsky}, \citenamefont {Gas},
		\citenamefont {Sawicki}, \citenamefont {Kriegner}, \citenamefont
		{Zub{\'{a}}{\v{c}}}, \citenamefont {Olejn{\'{i}}k}, \citenamefont
		{Nov{\'{a}}k}, \citenamefont {Jungwirth}, \citenamefont {Shahrokhvand},
		\citenamefont {Zeitler}, \citenamefont {Dhesi},\ and\ \citenamefont
		{Maccherozzi}}]{Wang2020}%
	\BibitemOpen
	\bibfield  {author} {\bibinfo {author} {\bibfnamefont {M.}~\bibnamefont
			{Wang}}, \bibinfo {author} {\bibfnamefont {C.}~\bibnamefont {Andrews}},
		\bibinfo {author} {\bibfnamefont {S.}~\bibnamefont {Reimers}}, \bibinfo
		{author} {\bibfnamefont {O.~J.}\ \bibnamefont {Amin}}, \bibinfo {author}
		{\bibfnamefont {P.}~\bibnamefont {Wadley}}, \bibinfo {author} {\bibfnamefont
			{R.~P.}\ \bibnamefont {Campion}}, \bibinfo {author} {\bibfnamefont {S.~F.}\
			\bibnamefont {Poole}}, \bibinfo {author} {\bibfnamefont {J.}~\bibnamefont
			{Felton}}, \bibinfo {author} {\bibfnamefont {K.~W.}\ \bibnamefont {Edmonds}},
		\bibinfo {author} {\bibfnamefont {B.~L.}\ \bibnamefont {Gallagher}}, \bibinfo
		{author} {\bibfnamefont {A.~W.}\ \bibnamefont {Rushforth}}, \bibinfo {author}
		{\bibfnamefont {O.}~\bibnamefont {Makarovsky}}, \bibinfo {author}
		{\bibfnamefont {K.}~\bibnamefont {Gas}}, \bibinfo {author} {\bibfnamefont
			{M.}~\bibnamefont {Sawicki}}, \bibinfo {author} {\bibfnamefont
			{D.}~\bibnamefont {Kriegner}}, \bibinfo {author} {\bibfnamefont
			{J.}~\bibnamefont {Zub{\'{a}}{\v{c}}}}, \bibinfo {author} {\bibfnamefont
			{K.}~\bibnamefont {Olejn{\'{i}}k}}, \bibinfo {author} {\bibfnamefont
			{V.}~\bibnamefont {Nov{\'{a}}k}}, \bibinfo {author} {\bibfnamefont
			{T.}~\bibnamefont {Jungwirth}}, \bibinfo {author} {\bibfnamefont
			{M.}~\bibnamefont {Shahrokhvand}}, \bibinfo {author} {\bibfnamefont
			{U.}~\bibnamefont {Zeitler}}, \bibinfo {author} {\bibfnamefont {S.~S.}\
			\bibnamefont {Dhesi}},\ and\ \bibinfo {author} {\bibfnamefont
			{F.}~\bibnamefont {Maccherozzi}},\ }\href
	{https://doi.org/10.1103/PhysRevB.101.094429} {\bibfield  {journal} {\bibinfo
			{journal} {Phys. Rev. B}\ }\textbf {\bibinfo {volume} {101}},\ \bibinfo
		{pages} {094429} (\bibinfo {year} {2020})}\BibitemShut {NoStop}%
	\bibitem [{\citenamefont {Fink}(1964)}]{Fink1964}%
	\BibitemOpen
	\bibfield  {author} {\bibinfo {author} {\bibfnamefont {H.~J.}\ \bibnamefont
			{Fink}},\ }\href {https://doi.org/10.1103/PhysRev.133.A1322} {\bibfield
		{journal} {\bibinfo  {journal} {Phys. Rev.}\ }\textbf {\bibinfo {volume}
			{133}},\ \bibinfo {pages} {A1322} (\bibinfo {year} {1964})}\BibitemShut
	{NoStop}%
	\bibitem [{\citenamefont {Searle}\ and\ \citenamefont
		{Wang}(1968)}]{Searle1968}%
	\BibitemOpen
	\bibfield  {author} {\bibinfo {author} {\bibfnamefont {C.~W.}\ \bibnamefont
			{Searle}}\ and\ \bibinfo {author} {\bibfnamefont {S.~T.}\ \bibnamefont
			{Wang}},\ }\href {https://doi.org/10.1063/1.1656155} {\bibfield  {journal}
		{\bibinfo  {journal} {J. Appl. Phys.}\ }\textbf {\bibinfo {volume} {39}},\
		\bibinfo {pages} {1025} (\bibinfo {year} {1968})}\BibitemShut {NoStop}%
	\bibitem [{\citenamefont {Moriyama}\ \emph {et~al.}(2019)\citenamefont
		{Moriyama}, \citenamefont {Hayashi}, \citenamefont {Yamada}, \citenamefont
		{Shima}, \citenamefont {Ohya},\ and\ \citenamefont {Ono}}]{Moriyama2019}%
	\BibitemOpen
	\bibfield  {author} {\bibinfo {author} {\bibfnamefont {T.}~\bibnamefont
			{Moriyama}}, \bibinfo {author} {\bibfnamefont {K.}~\bibnamefont {Hayashi}},
		\bibinfo {author} {\bibfnamefont {K.}~\bibnamefont {Yamada}}, \bibinfo
		{author} {\bibfnamefont {M.}~\bibnamefont {Shima}}, \bibinfo {author}
		{\bibfnamefont {Y.}~\bibnamefont {Ohya}},\ and\ \bibinfo {author}
		{\bibfnamefont {T.}~\bibnamefont {Ono}},\ }\href
	{https://doi.org/10.1103/PhysRevMaterials.3.051402} {\bibfield  {journal}
		{\bibinfo  {journal} {Phys. Rev. Mater.}\ }\textbf {\bibinfo {volume} {3}},\
		\bibinfo {pages} {051402} (\bibinfo {year} {2019})}\BibitemShut {NoStop}%
	\bibitem [{\citenamefont {Lebrun}\ \emph {et~al.}(2020)\citenamefont {Lebrun},
		\citenamefont {Ross}, \citenamefont {Gomonay}, \citenamefont {Baltz},
		\citenamefont {Ebels}, \citenamefont {Barra}, \citenamefont {Qaiumzadeh},
		\citenamefont {Brataas}, \citenamefont {Sinova},\ and\ \citenamefont
		{Kl{\"{a}}ui}}]{Lebrun2020}%
	\BibitemOpen
	\bibfield  {author} {\bibinfo {author} {\bibfnamefont {R.}~\bibnamefont
			{Lebrun}}, \bibinfo {author} {\bibfnamefont {A.}~\bibnamefont {Ross}},
		\bibinfo {author} {\bibfnamefont {O.}~\bibnamefont {Gomonay}}, \bibinfo
		{author} {\bibfnamefont {V.}~\bibnamefont {Baltz}}, \bibinfo {author}
		{\bibfnamefont {U.}~\bibnamefont {Ebels}}, \bibinfo {author} {\bibfnamefont
			{A.~L.}\ \bibnamefont {Barra}}, \bibinfo {author} {\bibfnamefont
			{A.}~\bibnamefont {Qaiumzadeh}}, \bibinfo {author} {\bibfnamefont
			{A.}~\bibnamefont {Brataas}}, \bibinfo {author} {\bibfnamefont
			{J.}~\bibnamefont {Sinova}},\ and\ \bibinfo {author} {\bibfnamefont
			{M.}~\bibnamefont {Kl{\"{a}}ui}},\ }\href
	{https://doi.org/10.1038/s41467-020-20155-7} {\bibfield  {journal} {\bibinfo
			{journal} {Nat. Commun.}\ }\textbf {\bibinfo {volume} {11}},\ \bibinfo
		{pages} {1} (\bibinfo {year} {2020})}\BibitemShut {NoStop}%
	\bibitem [{\citenamefont {Hamdi}\ \emph {et~al.}(2022)\citenamefont {Hamdi},
		\citenamefont {Posva},\ and\ \citenamefont {Grundler}}]{Hamdi2022a}%
	\BibitemOpen
	\bibfield  {author} {\bibinfo {author} {\bibfnamefont {M.}~\bibnamefont
			{Hamdi}}, \bibinfo {author} {\bibfnamefont {F.}~\bibnamefont {Posva}},\ and\
		\bibinfo {author} {\bibfnamefont {D.}~\bibnamefont {Grundler}}\ }\href
	{https://doi.org/10.48550/arxiv.2212.11887} {10.48550/arxiv.2212.11887}
	(\bibinfo {year} {2022})\BibitemShut {NoStop}%
	\bibitem [{\citenamefont {Zhou}\ \emph {et~al.}(2022)\citenamefont {Zhou},
		\citenamefont {Liao}, \citenamefont {Guo}, \citenamefont {Bai}, \citenamefont
		{Zhao}, \citenamefont {Wan}, \citenamefont {Huang}, \citenamefont {Han},
		\citenamefont {Qiao}, \citenamefont {You}, \citenamefont {Chen},
		\citenamefont {Chen}, \citenamefont {Zhou}, \citenamefont {Han},
		\citenamefont {Pan},\ and\ \citenamefont {Song}}]{Zhou2022}%
	\BibitemOpen
	\bibfield  {author} {\bibinfo {author} {\bibfnamefont {Y.}~\bibnamefont
			{Zhou}}, \bibinfo {author} {\bibfnamefont {L.}~\bibnamefont {Liao}}, \bibinfo
		{author} {\bibfnamefont {T.}~\bibnamefont {Guo}}, \bibinfo {author}
		{\bibfnamefont {H.}~\bibnamefont {Bai}}, \bibinfo {author} {\bibfnamefont
			{M.}~\bibnamefont {Zhao}}, \bibinfo {author} {\bibfnamefont {C.}~\bibnamefont
			{Wan}}, \bibinfo {author} {\bibfnamefont {L.}~\bibnamefont {Huang}}, \bibinfo
		{author} {\bibfnamefont {L.}~\bibnamefont {Han}}, \bibinfo {author}
		{\bibfnamefont {L.}~\bibnamefont {Qiao}}, \bibinfo {author} {\bibfnamefont
			{Y.}~\bibnamefont {You}}, \bibinfo {author} {\bibfnamefont {C.}~\bibnamefont
			{Chen}}, \bibinfo {author} {\bibfnamefont {R.}~\bibnamefont {Chen}}, \bibinfo
		{author} {\bibfnamefont {Z.}~\bibnamefont {Zhou}}, \bibinfo {author}
		{\bibfnamefont {X.}~\bibnamefont {Han}}, \bibinfo {author} {\bibfnamefont
			{F.}~\bibnamefont {Pan}},\ and\ \bibinfo {author} {\bibfnamefont
			{C.}~\bibnamefont {Song}},\ }\href
	{https://doi.org/10.1038/s41467-022-31531-w} {\bibfield  {journal} {\bibinfo
			{journal} {Nat. Commun.}\ }\textbf {\bibinfo {volume} {13}},\ \bibinfo
		{pages} {3723} (\bibinfo {year} {2022})}\BibitemShut {NoStop}%
	\bibitem [{\citenamefont {Li}\ \emph {et~al.}(2020{\natexlab{b}})\citenamefont
		{Li}, \citenamefont {Shen}, \citenamefont {Bai}, \citenamefont {Wang},
		\citenamefont {Zhang}, \citenamefont {Xia}, \citenamefont {Ezawa},
		\citenamefont {Tretiakov}, \citenamefont {Xu}, \citenamefont {Mruczkiewicz},
		\citenamefont {Krawczyk}, \citenamefont {Xu}, \citenamefont {Evans},
		\citenamefont {Chantrell},\ and\ \citenamefont {Zhou}}]{Li2020a}%
	\BibitemOpen
	\bibfield  {author} {\bibinfo {author} {\bibfnamefont {X.}~\bibnamefont
			{Li}}, \bibinfo {author} {\bibfnamefont {L.}~\bibnamefont {Shen}}, \bibinfo
		{author} {\bibfnamefont {Y.}~\bibnamefont {Bai}}, \bibinfo {author}
		{\bibfnamefont {J.}~\bibnamefont {Wang}}, \bibinfo {author} {\bibfnamefont
			{X.}~\bibnamefont {Zhang}}, \bibinfo {author} {\bibfnamefont
			{J.}~\bibnamefont {Xia}}, \bibinfo {author} {\bibfnamefont {M.}~\bibnamefont
			{Ezawa}}, \bibinfo {author} {\bibfnamefont {O.~A.}\ \bibnamefont
			{Tretiakov}}, \bibinfo {author} {\bibfnamefont {X.}~\bibnamefont {Xu}},
		\bibinfo {author} {\bibfnamefont {M.}~\bibnamefont {Mruczkiewicz}}, \bibinfo
		{author} {\bibfnamefont {M.}~\bibnamefont {Krawczyk}}, \bibinfo {author}
		{\bibfnamefont {Y.}~\bibnamefont {Xu}}, \bibinfo {author} {\bibfnamefont
			{R.~F.}\ \bibnamefont {Evans}}, \bibinfo {author} {\bibfnamefont {R.~W.}\
			\bibnamefont {Chantrell}},\ and\ \bibinfo {author} {\bibfnamefont
			{Y.}~\bibnamefont {Zhou}},\ }\href
	{https://doi.org/10.1038/s41524-020-00435-y} {\bibfield  {journal} {\bibinfo
			{journal} {npj Comput. Mater.}\ }\textbf {\bibinfo {volume} {6}},\ \bibinfo
		{pages} {1} (\bibinfo {year} {2020}{\natexlab{b}})}\BibitemShut {NoStop}%
	\bibitem [{\citenamefont {Lepadatu}(2020)}]{Lepadatu2020}%
	\BibitemOpen
	\bibfield  {author} {\bibinfo {author} {\bibfnamefont {S.}~\bibnamefont
			{Lepadatu}},\ }\href {https://doi.org/10.1063/5.0024382} {\bibfield
		{journal} {\bibinfo  {journal} {J. Appl. Phys.}\ }\textbf {\bibinfo {volume}
			{128}},\ \bibinfo {pages} {243902} (\bibinfo {year} {2020})}\BibitemShut
	{NoStop}%
	\bibitem [{\citenamefont {Crameri}()}]{Crameri2021}%
	\BibitemOpen
	\bibfield  {author} {\bibinfo {author} {\bibfnamefont {F.}~\bibnamefont
			{Crameri}},\ }\bibfield  {journal} {\bibinfo  {journal} {Zenodo}\ }\href
	{https://doi.org/10.5281/ZENODO.5501399} {10.5281/ZENODO.5501399}\BibitemShut
	{NoStop}%
	\bibitem [{\citenamefont {Crameri}\ \emph {et~al.}(2020)\citenamefont
		{Crameri}, \citenamefont {Shephard},\ and\ \citenamefont
		{Heron}}]{Crameri2020}%
	\BibitemOpen
	\bibfield  {author} {\bibinfo {author} {\bibfnamefont {F.}~\bibnamefont
			{Crameri}}, \bibinfo {author} {\bibfnamefont {G.~E.}\ \bibnamefont
			{Shephard}},\ and\ \bibinfo {author} {\bibfnamefont {P.~J.}\ \bibnamefont
			{Heron}},\ }\href {https://doi.org/10.1038/s41467-020-19160-7} {\bibfield
		{journal} {\bibinfo  {journal} {Nat. Commun.}\ }\textbf {\bibinfo {volume}
			{11}},\ \bibinfo {pages} {1} (\bibinfo {year} {2020})}\BibitemShut {NoStop}%
\end{thebibliography}

%

\end{document}